%% file: main.tex
\documentclass{selfevolagent}

\usepackage{times}
\usepackage{latexsym}
\usepackage{amsmath}
\usepackage{enumitem}

\usepackage[T1]{fontenc}

\usepackage[utf8]{inputenc}

\usepackage{microtype}

\usepackage{graphicx}

\usepackage{booktabs}
\usepackage{caption}
\usepackage{hyperref}
\usepackage{xurl}       %
\usepackage{lineno}
\usepackage{tabularx}

\usepackage{amsthm,amssymb}
\usepackage{bm}         %
\usepackage{siunitx}

\usepackage[dvipsnames,table]{xcolor}
\usepackage{algorithm}
\usepackage[noend]{algpseudocode}
\usepackage{multirow}
\usepackage{enumitem}
\usepackage{wrapfig}
\usepackage[most]{tcolorbox}

\usepackage{makecell}
\usepackage{pifont}
\usepackage{listings} 

\PassOptionsToPackage{capitalize,noabbrev}{cleveref}
\input{yaml_style}

\begin{document}

\title{\raisebox{-0.2\height}{\includegraphics[height=2.0em]{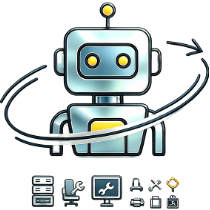}}\hspace{0.6em}OptiMAS: Automatically Optimize Multi-Agent System}
\author[1]{Yuxin~Cheng\cofirst}
\author[2]{Chang~Liu\cofirst\corrauthor}
\author[3]{Hanxin~Yu}
\author[2]{Haochen~Tan}
\author[1]{Taiqiang~Wu}
\author[5]{Weiqiang~Jin}
\author[2]{Jie~Ran}
\author[4]{Kaibo~Wang}
\author[2]{Xiaoguang~Li}
\author[2]{Haoli~Bai}
\author[1]{Graziano~Chesi}
\author[1]{Ngai~Wong\corrauthor}
\affiliation[1]{The University of Hong Kong}
\affiliation[2]{Huawei Foundation Model Department, Huawei Technologies Ltd.}
\affiliation[3]{City University of Hong Kong}
\affiliation[4]{The Hong Kong University of Science and Technology}
\affiliation[5]{Xi'an Jiaotong University}
\code{\url{https://github.com/Opti-MAS/OptiMAS}}

\abstract{
Automated evolution of Multi-Agent Systems (MAS) holds significant potential for reducing the manual effort required to design and optimize LLM-based agent architectures. However, extant search-based paradigms face a fundamental trade-off, where an expanded optimization scope exacerbates evolutionary instability, while discrete branch-and-discard search isolates insights across lineages. To address these limitations, we propose a continuous, data-driven optimization paradigm built upon a unified ReAct-based infrastructure that reconciles a broad optimization scope with operational stability. Under this paradigm, we present \emph{OptiMAS}, a task-agnostic agentic optimizer that leverages textual interaction trajectories and task feedback as loss signals for end-to-end MAS evolution. Equipped with a novel dual-track memory mechanism, OptiMAS sustains performance improvement over extended optimization horizons. Evaluation on four heterogeneous agentic benchmarks with three varying scale and accessibility LLM backbones, demonstrates that OptiMAS consistently achieves competitive or superior accuracy relative to both domain-specialized hand-crafted systems and existing evolutionary methods. Our work establishes a practical milestone toward robust, automated MAS evolution.
}

\makeatletter
\renewcommand{\authorlist}{%
  \raggedright
  Yuxin~Cheng\cofirst$^1$,\ \  Chang~Liu\cofirst\corrauthor$^2$,\ \  Hanxin~Yu$^3$ \\[0.15cm]
  Haochen~Tan$^2$,\ \  Taiqiang~Wu$^1$,\ \  Weiqiang~Jin$^5$,\ \  Jie~Ran$^2$,\ \  Kaibo~Wang$^4$ \\[0.15cm]
  Xiaoguang~Li$^2$,\ \  Haoli~Bai$^2$,\ \  Graziano~Chesi$^1$,\ \  Ngai~Wong\corrauthor$^1$\par
}
\makeatother

\maketitle
\begingroup
\renewcommand{\thefootnote}{\fnsymbol{footnote}}
\footnotetext[2]{Co-first authors.}
\footnotetext[3]{Corresponding authors.}
\endgroup

\input{section/1_introduction}
\input{section/2_relatedworks}
\input{section/3_method}
\input{section/4_experiments}
\input{section/5_conclusion}

\input{section/7_limitations}

\bibliography{my}
\bibliographystyle{plainnat}

\input{main.bbl}
\input{section/8_appendix}

\end{document}

%% file: yaml_style.tex
\definecolor{techDark}{rgb}{0.08, 0.09, 0.12}       
\definecolor{mdHeader}{rgb}{0.03, 0.27, 0.49}       
\definecolor{codeInline}{rgb}{0.7, 0.0, 0.13}       
\definecolor{nordGray}{rgb}{0.45, 0.50, 0.58}       
\definecolor{techBg}{rgb}{0.95, 0.96, 0.97}         
\definecolor{borderGray}{rgb}{0.80, 0.83, 0.87}     

\lstdefinestyle{agentStyle}{
  basicstyle=\ttfamily\color{techDark},             
  columns=fullflexible,
  keepspaces=true,
  breaklines=true,
  showstringspaces=false,
  tabsize=2,
  commentstyle=\color{nordGray},
  stringstyle=\color{codeInline},
  keywordstyle=\color{mdHeader}\bfseries,
  keywords={pytest,git,python,Docker,container,workspace,name,description,agents,topology}
}

\newtcblisting{yamlbox}{
  listing only,
  breakable,
  colback=techBg,                                   
  colframe=borderGray,                              
  boxrule=0.6pt,
  arc=3pt,                                          
  left=10pt,right=10pt,top=10pt,bottom=10pt,        
  listing options={
    style=agentStyle,
    basicstyle=\ttfamily\footnotesize\color{techDark},
    breaklines=true,        
    breakindent=0pt,        
    breakautoindent=false   
  }
}

\newtcolorbox{treebox}{
  breakable,
  colback=techBg,                                   
  colframe=borderGray,                              
  boxrule=0.6pt,
  arc=3pt,                                          
  left=10pt,right=10pt,top=10pt,bottom=10pt,
  fontupper=\ttfamily\footnotesize\color{techDark} 
}

\newtcolorbox{promptbox}{
  breakable,
  colback=techBg,                                    
  colframe=borderGray,                              
  boxrule=0.6pt,
  arc=3pt,                                           
  left=10pt,right=10pt,top=10pt,bottom=10pt,
  before=\noindent,
  fontupper=\ttfamily\footnotesize\color{techDark} 
}

%% file: section/1_introduction.tex
\section{Introduction}
\label{sec:intro}

\input{section/figure/paradigm}
Collaborative multi-agent systems (MAS) have fundamentally expanded the operational envelope of large language models (LLMs), extending their reach from advanced cognitive reasoning to pragmatic real-world information foraging~\cite{wei2025browsecomp,mialon2023gaia,liu2026uis} and autonomous software engineering~\cite{swebenchverified, yang2024swe,jimenez2024swe}. However, manual MAS construction remains inherently labor-intensive, demanding meticulous prompt calibration, capability augmentation, and orchestration design through iterative empirical trials. While recent work~\cite{hu2024automated} advocates for automated agent evolution, this nascent field has yet to match the efficacy and reliability of carefully handcrafted systems in practical deployments.

Existing automated MAS evolution efforts center on two axes: \emph{optimization scope} and \emph{evolutionary paradigm}~\cite{gao2025survey}. The optimization scope governs which facets of a MAS undergo evolution, ranging from prompt tuning to autonomous agent creation~\cite{shinn2023reflexion, sun2023adaplanner, khattab2023dspy, fernando2023promptbreeder, yuan2024evoagent, wang2023voyager, qiu2025alita, hu2024self}. Recent full-stack approaches that adapt the agent infrastructure itself at test time~\cite{zhang2025darwin} have further automated the design pipeline, progressively minimizing human intervention. On the paradigm axis, search-based methods remain dominant~\cite{zhang2024aflow,zhang2025evoflow,hu2026evolutionary}. As illustrated in~\Cref{fig:paradigm}, evolved MAS candidates are compiled into a structured pool, with subsequent generations inheriting from and expanding upon diverging ancestral branches~\cite{lee2026meta, zhang2025maas}. Despite these advances, handcrafted MAS architectures continue to prevail in industrial production~\cite{xia2025livesweagent}, as current automated approaches face a persistent dilemma: (1)~broadening the optimization scope exacerbates evolutionary instability, impeding robust generalization across diverse tasks~\cite{zhang2024aflow, zhang2025evoflow}; and (2)~search-oriented mechanisms encounter severe performance bottlenecks as accumulated evolutionary insights remain confined to isolated lineages and cannot transfer across branches.

To overcome these structural limitations, we propose \textbf{OptiMAS}, an optimization-based MAS evolution paradigm driven by a general-purpose agentic optimizer. We first establish a ReAct-centric~\cite{yao2022react} foundational infrastructure shared symmetrically by the evolving MAS and the optimizer. This unified substrate instantiates minimal configurations into robust, operational multi-agent systems, effectively bridging the gap between an expanded optimization scope and evolutionary stability~\cite{zhang2025darwin}.
In contrast to search-based methods, our framework cultivates a MAS from inception through a data-driven optimization pipeline. As illustrated in~\Cref{fig:paradigm}, the evolving MAS executes tasks on a partitioned \emph{train set} at each step. Its complete action trajectories and performance metrics are then framed as a \emph{loss} signal and routed to the optimizer for textual \emph{gradient} induction and structural refinement. To preempt overfitting, generalization is continuously calibrated via execution accuracy on an optimizer-invisible \emph{validation} set, with a held-out \emph{test} set reserved solely for final evaluation. To sustain coherent optimization across an extended evolutionary horizon, OptiMAS incorporates a \emph{dual-track memory mechanism} that continuously accumulates, validates, and retrieves evolutionary insights. Extensive experiments across heterogeneous benchmarks~\cite{wei2025browsecomp, swebenchverified, mialon2023gaia, styles2024workbench} demonstrate the exceptional efficacy of our framework. Our principal contributions are:
\begin{itemize}
\item We introduce an optimization-based MAS evolution paradigm underpinned by a unified infrastructure that stabilizes the evolution while accommodating broad-scope MAS design.
\item We present OptiMAS, a task-agnostic agentic optimizer featuring dual-track memory mechanism that enables steady performance compounding throughout prolonged optimization.
\item We provide comprehensive evaluation on four challenging benchmarks, demonstrating the substantial potential of our framework for synthesizing practically deployable MAS.
\end{itemize}

%% file: section/figure/paradigm.tex
\begin{figure*}
    \centering
    \includegraphics[width=\linewidth]{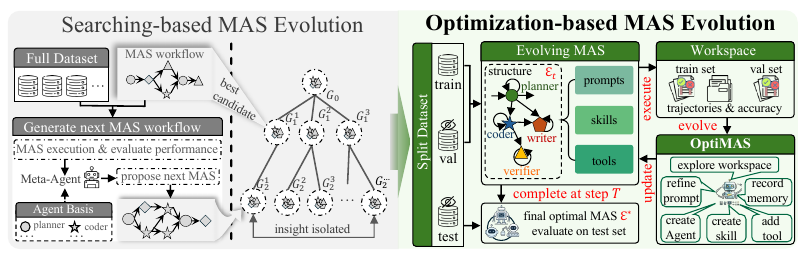}
    \caption{\textbf{Comparison of MAS evolution paradigms.} \textbf{(Left)} Search-based methods frame evolution as stochastic tree-branching ($G_0 \!\to\! G_t^i$) over full datasets, incurring low efficiency with insights isolated across branches.  \textbf{(Right)} OptiMAS  reformulates evolution as a data-driven continuous optimization over partitioned data (train/val/test), channeling action trajectories and accuracy as loss signals for OptiMAS to derive textual gradients and perform end-to-end MAS updates, yielding stable, sample-efficient, and overfit-resistant synthesis.}
    \label{fig:paradigm}
\end{figure*}

%% file: section/2_relatedworks.tex
\section{Related Works}
\label{sec:relat}

\subsection{LLM-based Agent and MAS}
Chain-of-thought prompting~\cite{wei2022chain} first endowed LLMs with multi-step reasoning, and ReAct~\cite{yao2022react} closed the perception-action loop by interleaving reasoning traces with grounded tool calls, establishing a paradigm that underpins most contemporary agent architectures. Subsequent advances in verbal self-reflection~\cite{shinn2023reflexion}, autonomous tool acquisition~\cite{schick2023toolformer}, and task-level orchestration~\cite{shen2023hugginggpt} further broadened the single-agent design space. Recognizing that complex objectives benefit from division of labor, researchers organized multiple agents into collaborative MAS. CAMEL~\cite{li2023camel} and AutoGen~\cite{wu2024autogen} explored role-play and multi-turn coordination. MetaGPT~\cite{hong2023metagpt} and ChatDev~\cite{qian2024chatdev} imposed structured workflows for software development. AgentVerse~\cite{chen2024agentverse} and multi-agent debate~\cite{du2023improving} demonstrated that dynamic team assembly and adversarial interaction improve both performance and factual consistency. Despite this architectural diversity, existing MAS remain hand-designed, task-specific, and static once deployed~\cite{zhou2025multi, wang2023lego}, motivating the pursuit of automated, self-evolving MAS design.
 
\subsection{Automated MAS Evolution}
Automated MAS design spans a widening spectrum of design freedom~\cite{gao2025survey}. Prompt-level methods (DSPy~\cite{khattab2023dspy}, PromptBreeder~\cite{fernando2023promptbreeder}) and agent-profiling approaches (EvoAgent~\cite{yuan2024evoagent}) optimize within frozen topologies. Workflow-level search, represented by ADAS~\cite{hu2024automated}, AFlow~\cite{zhang2024aflow}, GPTSwarm~\cite{zhuge2024gptswarm}, and AgentSquare~\cite{shang2025agentsquare}, extends evolution to control flow, yet the resulting structures typically chain shallow, single-call agents, discard all non-selected search branches, and draw from a fixed role vocabulary. Query-adaptive extensions such as MaAS~\cite{zhang2025maas}, EvoFlow~\cite{zhang2025evoflow}, MAS-GPT~\cite{ye2025masgpt}, and FlowReasoner~\cite{gao2025flowreasoner} relax the static-workflow assumption but inherit the shallow-agent limitation while introducing substantial training overhead. At the opposite extreme, fully self-referential systems (G\"{o}del Agent~\cite{yin2025godelagentselfreferentialagent}, Darwin G\"{o}del Machine~\cite{zhang2025darwin}) evolve the agent codebase itself, yet presuppose strong code-generation backbones and frequently yield non-executable configurations. Our work navigates this tension by jointly evolving prompts, skills, orchestration topology, and tool configurations under a stable, protocol-guided optimization framework in which each agent operates as a recursive ReAct-based reasoner with proactive inter-agent invocation, achieving a broader optimization scope than constrained workflow search while maintaining greater reliability than unconstrained codebase mutation.

%% file: section/3_method.tex
\section{Methodology}
\label{sec:metho}
 
\subsection{Infrastructure}
\label{subsec:backg}
We formalize the multi-agent infrastructure and delineate the optimization scope of OptiMAS.
 
\noindent\textbf{ReAct Agent.}
Our infrastructure adopts ReAct~\cite{yao2022react} as the atomic execution primitive for every agent. An agent $\mathcal{A}$ iteratively generates a thought $\tau$ followed by an action $a$, each conditioned on the accumulated context and the latest observation $o$ from environment $\Omega$. This interleaved history forms a trajectory $S_i = (\mathcal{P},\, \tau_1, a_1, o_1, \dots, \tau_i, a_i, o_i)$, where $\mathcal{P}$ denotes the initial prompt. At step $i$, the policy $\pi_{\mathcal{M}}$, parameterized by LLM backbone $\mathcal{M}$, produces:
\begin{equation}
\tau_{i+1}, a_{i+1}\!=\!\pi_{\mathcal{M}}(\cdot \mid S_i), \ \tau_{1}, a_{1}\!=\!\pi_{\mathcal{M}}(\cdot \mid \mathcal{P}),
\end{equation}
iterating until a capped horizon $n$ or an explicit termination signal.
 
\paragraph{Proactive Assign-Deliver.}
To compose ReAct agents into a collaborative MAS without hardcoded workflows, we adopt a \emph{proactive assign-deliver} protocol. Task delegation from a parent agent $\mathcal{A}$ to its sub-agents is modeled as a tool-invocation action. The sub-agent resolves the assignment and returns its solution to the parent context as a structured observation. Agents are thus activated on demand rather than by a fixed schedule, providing both flexibility and scalability required for non-deterministic, complex workflows.
 
\paragraph{Comprehensive Optimization Scope.}
This unified infrastructure yields a configuration-driven MAS whose optimization scope $\mathcal{E}$ spans from prompt tuning to the macroscopic orchestration graph $\mathcal{G}(\mathbb{A})$ over an evolving agent population $\mathbb{A}$:
\begin{equation}
    \mathcal{E} = \left\{ \mathbb{A},\, \mathcal{G}(\mathbb{A}),\, \left( \mathcal{M}_\mathcal{A},\, \mathcal{P}_\mathcal{A},\, \mathcal{K}_{\mathcal{A}},\, \mathcal{T}_\mathcal{A} \right)_{\mathcal{A} \in \mathbb{A}} \right\},
\end{equation}
where $\mathcal{M}_\mathcal{A}$, $\mathcal{P}_\mathcal{A}$, $\mathcal{K}_{\mathcal{A}}$, and $\mathcal{T}_\mathcal{A} \subseteq \mathbb{T}$ denote the LLM backbone, prompts, skill set, and accessible toolkit for each agent $\mathcal{A}$. To preserve operational stability, OptiMAS composes tools from a predefined macro-library $\mathbb{T}$ rather than generating code from scratch (Appendix~\ref{sec:appendix_infrastructure}).
Unlike prior methods that search a codified flow within a fixed population or static graph~\cite{zhang2024aflow, hu2024automated, zhang2025maas}, OptiMAS freely instantiates heterogeneous agent roles via prompt compilation $\mathcal{P}_\mathcal{A}$ and coordinates $\mathcal{G}(\mathbb{A})$ through assign-deliver actions. The sole constraint on $\mathcal{G}(\mathbb{A})$ is the directed acyclic graph (DAG) property, guaranteeing deterministic termination.

\input{section/algorithm/adaptive_sampling_strategy}

\subsection{Optimization Paradigm}

Inspired by deep learning~\cite{lecun2015deep, MachineLearningI}, we reformulate MAS evolution as an optimization-based paradigm. Instead of discretely selecting promising candidates and creating offspring along diverging branches, OptiMAS continuously refines the MAS configuration guided by textual gradients, thereby seeking monotonic performance improvement.
 
As illustrated in~\Cref{fig:paradigm}, the task query set is first partitioned into $\mathcal{X}_\text{train}$, $\mathcal{X}_\text{val}$, and $\mathcal{X}_\text{test}$. At each training step $t$, a batch $\mathcal{B}_t$ of size $n$ is drawn from $\mathcal{X}_\text{train}$ via the adaptive sampling strategy described in~\Cref{alg:adaptive_sampling}, which proactively revisits previous failures while preserving exploration dynamics. The current MAS $\mathcal{E}_{t-1}$ then executes on both $\mathcal{B}_t$ and $\mathcal{X}_\text{val}$, and the resulting outputs are evaluated against ground-truth references, yielding accuracy scores $R_\text{train}$ and $R_\text{val}$.
 
In the back-propagation stage, the agentic optimizer produces textual gradients and evolves $\mathcal{E}_{t-1}$ accordingly by deeply inspecting the workspace $\mathcal{W}_\text{train}$, which preserves the complete trajectory $S$ generated by $\mathcal{E}_{t-1}$ for each query along with all intermediate artifacts, e.g., code, crawled documents, working notes (\Cref{fig:optimas}a). Equipped with versatile tools, the optimizer diagnoses root causes of failures and identifies recurring success patterns. Unlike structured backward passes that separate gradient computation from parameter update~\cite{lecun2015deep,yellamraju2024textgrad}, our optimizer autonomously performs gradient analysis and MAS configuration improvement in an end-to-end manner, preserving coherence of reasoning throughout. Notably, the workspace of $\mathcal{E}_{t-1}$ on $\mathcal{X}_\text{val}$ is screened from the optimizer, while only the scalar accuracy $R_\text{val}$ provided as a generalization signal to guard against overfitting.
 
This inference-evolution cycle repeats until the maximum step $T$ is reached. The full algorithmic flow is given in~\Cref{alg:mas-evolution}.

\input{section/algorithm/training_paradigm}
\input{section/figure/optimizer}

\subsection{OptiMAS}
 
To achieve effective continual evolution across diverse agentic tasks, we propose OptiMAS, a general-purpose agentic optimizer specialized for MAS evolution (\Cref{fig:optimas}d), equipped with a comprehensive toolkit and dedicated skills $\mathcal{K}$ encoding knowledge of the MAS infrastructure and task-agnostic optimization scope (Appendix~\ref{sec:appendix_optimas_design}).

Despite this self-contained design, long-horizon MAS evolution poses stability challenges. LLM amnesia and hallucination erode reasoning consistency as context grows, while textual gradients lack strict descent guarantees. We therefore introduce a dual-track memory mechanism, namely \emph{plan-oriented short-term memory} and \emph{hypothesis-driven long-term memory}, enabling OptiMAS to accumulate, verify, and propagate evolutionary insights across steps.

\noindent\textbf{Plan-oriented short-term memory.}
MAS evolution on challenging tasks demands OptiMAS to process million-token trajectories for textual gradient extraction, making intra-step consistency critical. After obtaining performance metrics $R_\text{train/val}$ and preliminarily exploring $\mathcal{W}_\text{train}$, OptiMAS creates an instructional plan that serves as a structural spine throughout the back-propagation stage, mitigating forgetting and hallucination-induced drift. To balance flexibility with discipline, plans are managed via four statuses (\texttt{waiting}, \texttt{executing}, \texttt{completed}, \texttt{dropped}) and four operations (\texttt{add}, \texttt{modify}, \texttt{review}, \texttt{delete}). Crucially, OptiMAS is forbidden from erasing content, and inapplicable instructions are instead transitioned to \texttt{dropped} with mandatory justification, preserving a complete audit trail that prevents hallucination-affected inconsistencies. A programmatic completeness check at step termination enforces that all instructions reach a terminal status, redirecting OptiMAS to resume unfinished items otherwise (\Cref{fig:optimas}c).

\noindent\textbf{Hypothesis-driven long-term memory.}
While short-term memory governs intra-step coherence, MAS evolution demands \emph{cross-step continuity}. Insights from one batch must be empirically verified over subsequent iterations. Without persistent memory, the optimizer risks repeating failed interventions, accidentally reverting effective improvements, or conflating stochastic noise with systematic deficiencies.

We address this via hypothesis-driven long-term memory, whose core is a closed-loop lifecycle with verifiable rubrics governing how evolutionary insights accumulate across steps (\Cref{fig:optimas}b). Each hypothesis is initialized as \textsc{Propose} with OptiMAS's bootstrapped attributes, including classification, trajectory observations, identified textual gradient, prescribed MAS modifications, and falsifiable expectations. The hypothesis then transitions through a principled lifecycle:

\begin{itemize}
    \item \textsc{Propose} $\to$ \textsc{Enact}: upon applying the modification to $\mathcal{E}_{t}$.
    \item \textsc{Enact} $\to$ \textsc{Validate}: when trajectory-level evidence, not accuracy alone, confirms the expected improvement.
    \item \textsc{Enact} $\to$ \textsc{Refute}: when analysis reveals negative effects, triggering reversion.
    \item \textsc{Enact} $\to$ \textsc{Suspend}/\textsc{Dormant}: upon replacement by a successor or insufficient verification scenarios.
\end{itemize}
 
At each step, OptiMAS re-evaluates all active hypotheses against $\mathcal{W}_\text{train}$ evidence and maintains an accumulating evidence log. Every modification to evolving $\mathcal{E}_{t}$ must reference a hypothesis identifier, ensuring traceability.
 
By coupling each MAS modification to a falsifiable, evidence-grounded hypothesis, hypothesis-driven long-term memory transforms MAS evolution from undirected search into a principled, self-correcting optimization process, analogous to how a research log disciplines scientific experiments.
 

%% file: section/algorithm/adaptive_sampling_strategy.tex
\begin{algorithm}[t]
\small
\caption{Adaptive Sampling and Priority Weighting}
\label{alg:adaptive_sampling}
\begin{algorithmic}[1]
\Require $\mathcal{X}_{\text{train}}, \mathcal{F}_{\text{prev}}, \mathbf{p}, n, \phi$
\Function{AdaptiveSample}{$\mathcal{X}_{\text{train}}, \mathcal{F}_{\text{prev}}, \mathbf{p}, n, \phi$}
    \State $n_1 \leftarrow \lfloor \phi \cdot n \rfloor$, \;\; $n_2 \leftarrow n - n_1$
    \If{$|\mathcal{F}_{\text{prev}}| \ge n_1$}
        \State $\mathcal{B}_{\text{fail}} \sim \text{Categorical}(\mathcal{F}_{\text{prev}}, \mathbf{p}, n_1)$
        \State $\mathcal{B}_{\text{gen}} \sim \text{Categorical}(\mathcal{X}_{\text{train}} \setminus \mathcal{B}_{\text{fail}} , \mathbf{p}, n_2)$ 
    \Else
       \State $\mathcal{B}_{\text{fail}} \leftarrow \mathcal{F}_{\text{prev}}$;\; $n_{2} \leftarrow n - |\mathcal{F}_{\text{prev}}|$
        \State $\mathcal{B}_{\text{gen}} \sim \text{Categorical}(\mathcal{X}_{\text{train}} \setminus \mathcal{B}_{\text{fail}}, \mathbf{p}, n_{2})$ 
    \EndIf
    \State \Return $\mathcal{B} \leftarrow \mathcal{B}_{\text{gen}} \cup \mathcal{B}_{\text{fail}}$
\EndFunction
\Require update factors $\alpha, \beta$, bounds $p_{\min}, p_{\max}$, interval $\tau$
\Function{UpdateWeights}{$\mathbf{p}, \mathcal{B}, R_{\text{train}}, t$}
    \For{each sample $x_i \in \mathcal{B}$}
        \If{$\text{is\_correct}(x_i \mid R_{\text{train}})$}
            \State $p_i \leftarrow \max(p_{\min}, p_i \cdot \alpha)$
        \Else
            \State $p_i \leftarrow \min(p_{\max}, p_i \cdot \beta)$
        \EndIf
    \EndFor
    \If{$t \bmod \tau = 0$} $\mathbf{p} \leftarrow \mathbf{1}_{|\mathcal{X}_{\text{train}}|}$ \EndIf
    \State \Return $\mathbf{p}$
\EndFunction
\end{algorithmic}
\end{algorithm}

%% file: section/algorithm/training_paradigm.tex
\begin{algorithm}[t]
\small
\caption{Continuous MAS Evolution Paradigm}
\label{alg:mas-evolution}
\begin{algorithmic}[1]
\Require Initial MAS $\mathcal{E}_0$; Task queries $\mathcal{X}$; Epochs $T$; Batch size $n$; Retry ratio $\phi$
\Ensure Optimized Multi-Agent System $\mathcal{E}^{*}$

\State Partition query set $\mathcal{X} \rightarrow \{\mathcal{X}_{\text{train}}, \mathcal{X}_{\text{val}}\}$
\State Initialize sampling priority $\mathbf{p} \leftarrow \mathbf{1}_{|\mathcal{X}_{\text{train}}|}$
\Comment{\textit{Uniform init}}
\State $\mathcal{F}_{\text{prev}} \leftarrow \varnothing$
\Comment{\textit{Failed queries from previous step}}
\For{$t = 1$ \textbf{to} $T$}
    \State $\mathcal{B}_t \leftarrow \textsc{AdaptiveSample}(\mathcal{X}_{\text{train}}, \mathcal{F}_{\text{prev}}, \mathbf{p}, n, \phi)$
    \State $R_{\text{train}}, \mathcal{W}_{\text{train}} \leftarrow \textsc{Infer\&Evaluate}(\mathcal{E}_{t-1},\, \mathcal{B}_t)$
    \State $R_{\text{val}}, \_  \leftarrow \textsc{Infer\&Evaluate}(\mathcal{E}_{t-1},\, \mathcal{X}_{\text{val}})$
    \State $\mathcal{E}_{t}, \nabla_{\text{text}} \leftarrow \text{OptiMAS}(\mathcal{E}_{t-1},\mathcal{W}_{\text{train}}, R_{\text{train}}, R_{\text{val}})$
    \State $\mathcal{F}_{\text{prev}} \leftarrow \{ x_i \in \mathcal{B}_t \mid \text{is\_incorrect}(x_i) \}$
    \State $\mathbf{p} \leftarrow \text{\textsc{UpdateWeights}}(\mathbf{p}, \mathcal{B}_t, R_{\text{train}}, t)$
\EndFor
\State $\mathcal{E}^{*} \leftarrow \mathcal{E}_T$
\State \Return $\mathcal{E}^{*}$
\end{algorithmic}
\end{algorithm}

%% file: section/figure/optimizer.tex
\begin{figure*}
    \centering
    \includegraphics[width=\linewidth]{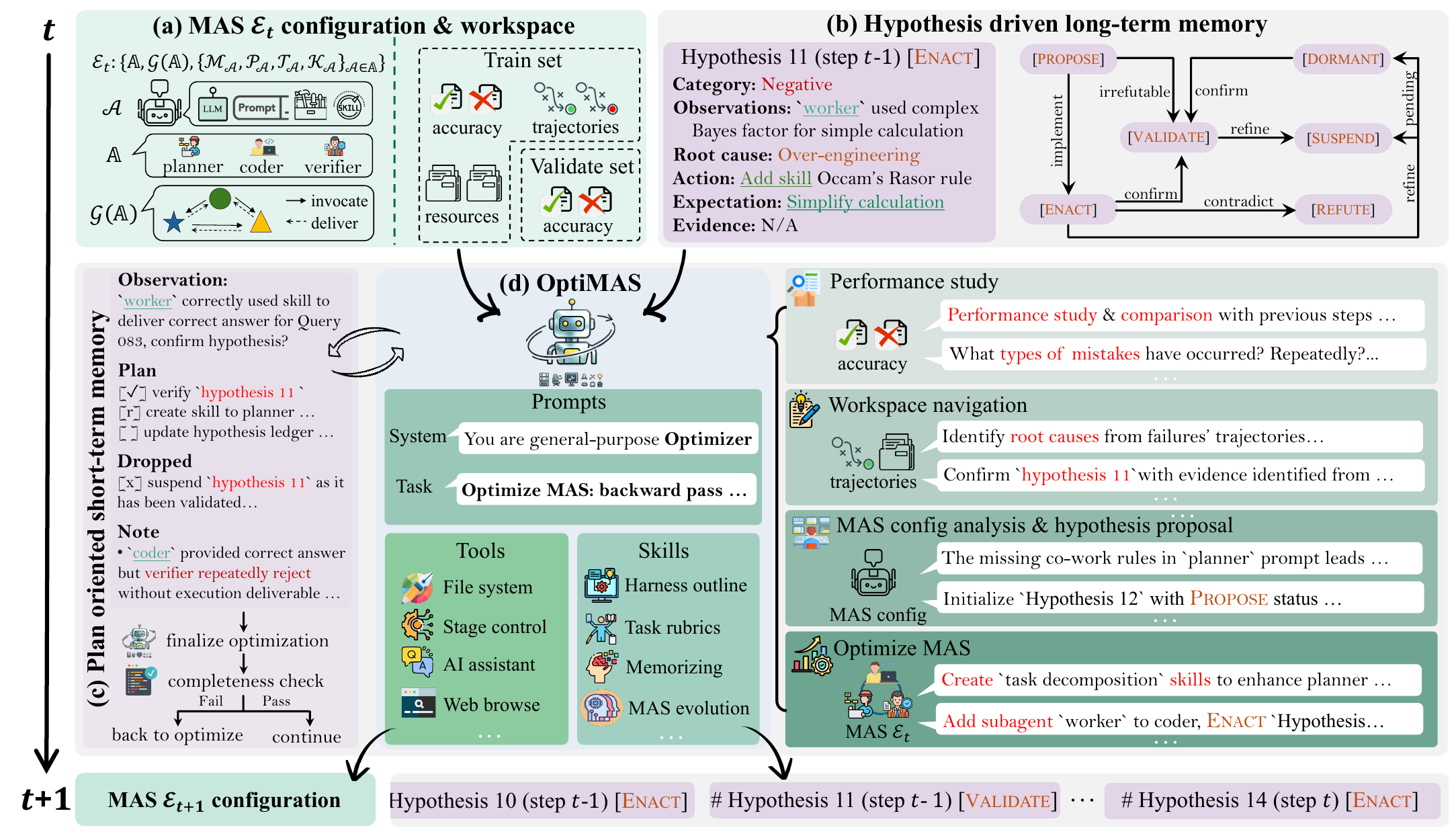}
    \caption{\textbf{OptiMAS optimizer architecture at step $t$.}  \textbf{(a)}~The MAS $\mathcal{E}_t$ executes on train/val queries, producing trajectories and accuracy. \textbf{(d)}~The optimizer diagnoses root causes via workspace inspection and evolves $\mathcal{E}_t$, governed by \textbf{(c)}~a plan-oriented short-term memory enforcing intra-step consistency and \textbf{(b)}~a hypothesis-driven long-term memory maintaining a state-tracked ledger of falsifiable hypotheses across steps, yielding $\mathcal{E}_{t+1}$.}
    \label{fig:optimas}
\end{figure*}

%% file: section/4_experiments.tex
\section{Experiments}
\label{sec:exper}

\input{section/table/main_results}

\subsection{Experiment Settings}
\label{sec:experiment_setting}
\noindent\textbf{Benchmarks.}
We evaluate on four benchmarks spanning heterogeneous agentic task settings to ensure comprehensive and rigorous assessment of evolutionary MAS approaches.
\emph{WorkBench}~\cite{styles2024workbench} tests multi-step planning, tool manipulation, and hierarchical execution on realistic workplace tasks.
\emph{GAIA-text}~\cite{kazemi2025big} poses complex reasoning questions requiring mathematical calculation, file parsing, and cross-tool orchestration.
\emph{BrowseComp}~\cite{wei2025browsecomp} targets long-horizon web navigation and programmatic multi-page information synthesis.
\emph{SWE-Bench-Verified (SWE-Bench)}~\cite{jimenez2023swe} tasks agents with resolving production-grade GitHub issues through multi-file code comprehension and repository modification.

To mitigate the prohibitive computational overhead while preserving evaluation integrity, we construct evaluation subsets of 100 queries each via reproducible seed-fixed random sampling for \emph{SWE-Bench-Verified} and \emph{BrowseComp}. For \emph{GAIA} and \emph{WorkBench}, we use the full set. Partition details of $\mathcal{X}_\text{train}$/$\mathcal{X}_\text{val}$/$\mathcal{X}_\text{test}$ are provided in Appendix~\ref{sec:appendix_benchmark_curation}. All baselines are evaluated on identical subsets and splits to ensure a strictly fair comparison.

\noindent\textbf{Metrics.}
We follow each benchmark's official metric protocol. \emph{SWE-Bench-Verified} and \emph{WorkBench} deliverables are verified via official sandboxed execution. \emph{BrowseComp} and \emph{GAIA-text} use a standardized LLM-as-a-Judge pipeline applied uniformly across all methods. All test results are averaged over three independent trials to account for LLM non-determinism.

\noindent\textbf{Baselines.}
We compare OptiMAS against two categories of competitive methodologies (details in Appendix~\ref{sec:appendix_baselines}):
\begin{itemize}[left=0pt]
    \item \textbf{Hand-crafted MAS}: domain-specialized systems including \emph{Tongyi-DeepResearch}~\cite{tongyidr} for \emph{GAIA-text}/\emph{BrowseComp} and \emph{SWE-Agent}~\cite{yang2024sweagent} for \emph{SWE-Bench-Verified}, alongside general-purpose \emph{SMoA}~\cite{li2025smoa} and \emph{Multi-Agent Debate}~\cite{du2023improving}.
    \item \textbf{Automated agent methods}: the state-of-the-art frameworks, including \emph{ADAS}~\cite{hu2024automated}, \emph{EvoAgent}~\cite{yuan2025evoagent}, and \emph{DGM}~\cite{zhang2025darwin}.
\end{itemize}
We additionally report the single-agent initialization $\mathcal{E}_0$, parameterized with minimalist prompts and basic toolkits, as a baseline isolating the intrinsic capacity of our infrastructure (Appendix~\ref{sec:appendix_mas_config}).

\noindent\textbf{Implementation Details.}
\emph{Gemini-3-Flash (Gemini3)} serves as the OptiMAS backbone and is mirrored as the meta-agent across all automated evolution baselines to ensure equitable comparison. Evolving MAS candidates are evaluated with three backbones: \emph{GPT-5-Nano (GPT5)}, \emph{Qwen-3.6-35B-A3B (Qwen3.6)}, and \emph{Gemini-3-Flash}. Hyperparameters and environment configurations are detailed in Appendix~\ref{sec:appendix_experiment_implementation}.

\subsection{Main Results}

Table~\ref{tab:main_results} reports task accuracy across four benchmarks and three backbone LLMs. We analyze the results along three axes: evolution efficacy, cross-domain generalization, and backbone universality.

\noindent\textbf{Evolution efficacy.}
OptiMAS achieves the highest accuracy in most benchmark--backbone configurations and improves over its single-agent initialization $\mathcal{E}_0$ in all conditions, with gains ranging from +2.8\% (BrowseComp/GPT5) to +45.4\% (WorkBench/GPT5). Notably, a single task-agnostic OptiMAS configuration is applied identically across all four benchmarks and three backbones without any task-specific adaptation, yet consistently yields positive evolution gains. We attribute this exceptional efficacy to the broad optimization scope and our hypothesis-driven memory mechanism, which enable OptiMAS to diagnose specific failure patterns and evolve targeted remedies regardless of the underlying domain. By contrast, the evolved systems produced by ADAS fall below our OptiMAS by up to 58.9\% on SWE-Bench/Gemini3, suggesting that evolving MAS along a continuous, evidence-grounded paradigm is more effective than discrete search for long-horizon agentic tasks that demand multi-step reasoning and complex coordination.

\input{section/figure/evolution_dynamics}

\noindent\textbf{Cross-domain generalization.}
OptiMAS matches or exceeds domain-specialized hand-crafted systems on their respective benchmarks while remaining fully transferable. On GAIA and BrowseComp, OptiMAS surpasses Tongyi-DR with both Qwen3.6 (81.2\%/43.8\% vs.\ 80.7\%/38.3\%) and Gemini3 (87.1\%/58.3\% vs.\ 75.8\%/48.3\%). On SWE-Bench, OptiMAS outperforms SWE-Agent across all three backbones. While these systems benefit from domain expertise, OptiMAS achieves competitive or superior performance on all four benchmarks from a general-purpose evolutionary optimizer, indicating that the breadth of the optimization scope enables task-adaptive capability.

\noindent\textbf{Backbone universality.}
OptiMAS delivers consistent improvements across backbone LLMs of diverse scales and accessibility, including the closed-source GPT-5-Nano (128k context), the open-source Qwen-3.6-35B-A3B (262k context), and the commercial frontier Gemini-3-Flash (1M context). The evolution gain is inversely correlated with backbone capacity, e.g., +45.4\%/+12.3\% on WorkBench/SWE-Bench under GPT5 versus +15.5\%/+4.4\% under Gemini3, indicating that evolved orchestration and skills effectively compensate for limited backbone reasoning. This universality across both proprietary and open-source models suggests that OptiMAS is applicable to practical deployment scenarios without constraints on the underlying LLM.

\subsection{Evolution Dynamics} 
Figure~\ref{fig:evolution_dynamics} traces the accuracy of each intermediate MAS $\mathcal{E}_t$ on the WorkBench train and test partitions over 10 optimization epochs. All three backbones exhibit progressive and near-monotonic improvement on both splits, with training accuracy gaining over +43\% (GPT5), +36\% (Qwen3.6), and +21\% (Gemini3) absolute, and test accuracy closely tracking these advances. The smooth ascending curves empirically validate the continuous optimization paradigm proposed in \Cref{sec:metho}. The hypothesis-driven long-term memory enables the optimizer to retain and compound insights across steps, producing sustained improvement that contrasts with the discrete, branch-and-discard dynamics of search-based methods. Equally notable is the narrow train-test gap maintained throughout evolution, indicating that the validation metric and adaptive sampling effectively prevent overfitting and ensure gains on $\mathcal{X}_\text{train}$ generalize to unseen queries. These results confirm that OptiMAS delivers stable, compounding MAS enhancement rather than volatile oscillation, a prerequisite for practical deployment.

\subsection{Ablation Analysis}
\input{section/figure/ablation_OptiMAS}

\noindent\textbf{Optimizer backbone.}
Figure~\ref{fig:ablation_optimas} compares the default Gemini-3-Flash OptiMAS against Qwen-3.6-35B-A3B across all four benchmarks and two MAS backbones. The Qwen-based OptiMAS yields positive gains over $\mathcal{E}_0$ in all eight configurations, confirming that the evolution framework is not contingent on a specific frontier model. The improvement is most pronounced on WorkBench (+8.5\% for Qwen-MAS, +38.3\% for GPT5-MAS) and SWE-Bench (+2.7\%/+6.1\%), where per-query agent working trajectories are relatively compact. However, the Qwen optimizer produces only marginal gains (+1.6\%/+1.1\% for Qwen-MAS) on GAIA and BrowseComp, while the Gemini optimizer achieves substantially larger improvements on the same configurations (+8.1\%/+9.9\%). This disparity is consistent with the context-length bottleneck: complex tasks involving long-horizon web navigation generate extensive workspace artifacts, degrading the quality of gradient diagnosis. These results suggest that optimizer context capacity acts as a scaling factor for evolution quality on trajectory-intensive tasks, while the underlying framework remains effective across optimizer scales.

\input{section/table/ablations}

\noindent\textbf{Component effectiveness.}
\Cref{tab:robustness_analysis} isolates each framework component on WorkBench and SWE-Bench. Removing hypothesis memory causes the most severe degradation, with SWE-Bench accuracy regressing by over $-$7\% on both backbones relative to $\mathcal{E}_0$. Without persistent inter-step memory, the optimizer cannot verify the effectiveness of prior interventions. Consequently, evolution reduces to isolated trial-and-error, which is particularly detrimental on complex tasks requiring sustained, coherent optimization. Removing adaptive sampling produces a similar pattern, with SWE-Bench performance falling below $\mathcal{E}_0$ on both backbones ($-$5.0\%/$-$0.5\%), as the optimizer loses the ability to examine previously failed queries and empirically validate its hypotheses. On the WorkBench, both ablations still yield gains over $\mathcal{E}_0$, indicating that the components become increasingly critical as task complexity grows. Reducing batch size degrades GPT5 on SWE-Bench by $-$5.0\%, because the weaker backbone frequently produces entirely incorrect batches that deprive the optimizer of positive signal for hypothesis validation. Removing the validation metric has a moderate effect overall but introduces slight overfitting on SWE-Bench/GPT5 ($-$0.5\%), consistent with its theoretical role for generalization.
 
\noindent\textbf{Cost and scalability.}
Statistically, OptiMAS processes 1.4M to 3.2M tokens per optimization step across four benchmarks, equivalent to approximately 750K to 2.4M words of technical content, with input constituting over 98\% of consumption. This volume, scaling with task-horizon complexity, reflects OptiMAS's ability to systematically explore large-scale workspaces containing extensive agent trajectories, intermediate code, retrieved documents, and working artifacts. Benefiting from the plan-oriented memory and advanced infrastructure, OptiMAS remains effective even when cumulative workspace content substantially exceeds the LLM's native context window, enabling thorough textual gradient analysis at a scale that meets industrial MAS evolution requirements. The evolved $\mathcal{E}^*$ is subsequently deployed at standard inference cost.

%% file: section/table/main_results.tex
\begin{table*}[t]
  \centering
  \caption{Main results (task accuracy, \%) across four benchmarks and three backbone LLMs: GPT5, Qwen3.6 and Gemini3 (abbreviated in column headers; full specifications in \cref{sec:experiment_setting}). \colorbox{green!15}{Green} and \colorbox{Violet!15}{purple} highlight the best and second-best results per backbone column. (--) denotes that such methodology is inapplicable or non-executable on given configurations. (*) marks evaluation subsets sampled via data curation(Appendix~\ref{sec:appendix_benchmark_curation}).}
  \label{tab:main_results}
  \scriptsize
  \setlength{\tabcolsep}{1.2pt}
  \renewcommand{\arraystretch}{1}
  \resizebox{\linewidth}{!}{
  \begin{tabular}{@{}lcccccccccccc@{}}
    \toprule
    \multirow{2}{*}{\textbf{Method}}
      & \multicolumn{3}{c}{\textbf{WorkBench (\%) $\uparrow$}}
      & \multicolumn{3}{c}{\textbf{GAIA (\%)$\uparrow$}}
      & \multicolumn{3}{c}{\textbf{BrowseComp* (\%)$\uparrow$}}
      & \multicolumn{3}{c}{\textbf{SWE-Bench* (\%)$\uparrow$}} \\
    \cmidrule(lr){2-4}\cmidrule(lr){5-7}\cmidrule(lr){8-10}\cmidrule(lr){11-13}
      & GPT5 & Qwen3.6 & Gemini3
      & GPT5 & Qwen3.6 & Gemini3
      & GPT5 & Qwen3.6 & Gemini3
      & GPT5 & Qwen3.6 & Gemini3 \\
    \midrule
        \rowcolor{gray!20} \multicolumn{13}{l}{\textit{Self-define Initial Agent}} \\ 
       Single $\mathcal{E}_0$ & 18.6 & \cellcolor{Violet!15} 65.6 & \cellcolor{Violet!15} 71.7 & 49.5 & 73.1 & \cellcolor{Violet!15} 80.7 & 8.3 & 33.9 & \cellcolor{Violet!15} 52.2 & \cellcolor{Violet!15} 29.4 & 61.1 & 72.8 \\
    \midrule
        \rowcolor{gray!20} \multicolumn{13}{l}{\textit{Hand-crafted MAS}} \\ 
      SMoA & 15.8 & 57.4 & 66.3 & 38.7 & 50.0 & 59.7 & 1.7 & 6.7 & 20.0 & -- & -- & -- \\
      MAS Debate & 15.0 & 39.2 & 44.2 & 32.3 & 38.7 & 50.0 & 5.0 & 11.7 & 20.0 & 13.3 &  23.3 & 26.7 \\
      SWE-Agent & -- & -- & -- & -- & -- & -- & -- & -- & -- & 16.7 & \cellcolor{Violet!15} 67.2 & \cellcolor{Violet!15} 75.0 \\
      Tongyi-DR & -- & -- & -- & \cellcolor{green!15} 62.9 & \cellcolor{Violet!15} 80.7 & 75.8 & \cellcolor{Violet!15} 10.0 & \cellcolor{Violet!15} 38.3 & 48.3 & -- & -- & -- \\
    \midrule
        \rowcolor{gray!20} \multicolumn{13}{l}{\textit{Evolutionary MAS}} \\ 
      ADAS & 41.6 & 50.8 & 62.1 & 35.5 & 27.4 & 45.2 & 3.3 & 8.3 & 13.3 & 18.3 & 21.7 & 18.3 \\
      EvoAgent & \cellcolor{Violet!15} 46.1 & 56.4 & 67.5 & 41.9 & 46.8 & 62.9 & 3.3 & 6.6 & 23.3 & 15.0 & 16.7 & 21.7  \\
      DGM & -- & -- & -- & -- & -- & -- & -- & -- & -- & 8.3 & 25.0 & 55.0 \\
      OptiMAS (Ours) & \cellcolor{green!15} 64.0\textsubscript{\textcolor{ForestGreen!80}{(+45.4)}} & \cellcolor{green!15} 84.5\textsubscript{\textcolor{ForestGreen!80}{(+18.9)}} & \cellcolor{green!15} 92.9\textsubscript{\textcolor{ForestGreen!80}{(+21.2)}} & \cellcolor{Violet!15} 56.9\textsubscript{\textcolor{ForestGreen!80}{(+7.4)}} & \cellcolor{green!15} 81.2\textsubscript{\textcolor{ForestGreen!80}{(+8.1)}} & \cellcolor{green!15} 87.1\textsubscript{\textcolor{ForestGreen!80}{(+6.4)}} & \cellcolor{green!15} 11.1\textsubscript{\textcolor{ForestGreen!80}{(+2.8)}} & \cellcolor{green!15} 43.8\textsubscript{\textcolor{ForestGreen!80}{(+9.9)}} & \cellcolor{green!15} 58.3\textsubscript{\textcolor{ForestGreen!80}{(+6.1)}} & \cellcolor{green!15} 41.7\textsubscript{\textcolor{ForestGreen!80}{(+12.3)}} & \cellcolor{green!15} 68.9\textsubscript{\textcolor{ForestGreen!80}{(+7.8)}} & \cellcolor{green!15} 77.2\textsubscript{\textcolor{ForestGreen!80}{(+4.4)}} \\
    \bottomrule
  \end{tabular}
  }
\end{table*}

%% file: section/figure/evolution_dynamics.tex
\begin{figure}
    \centering
    \includegraphics[width=0.7\linewidth]{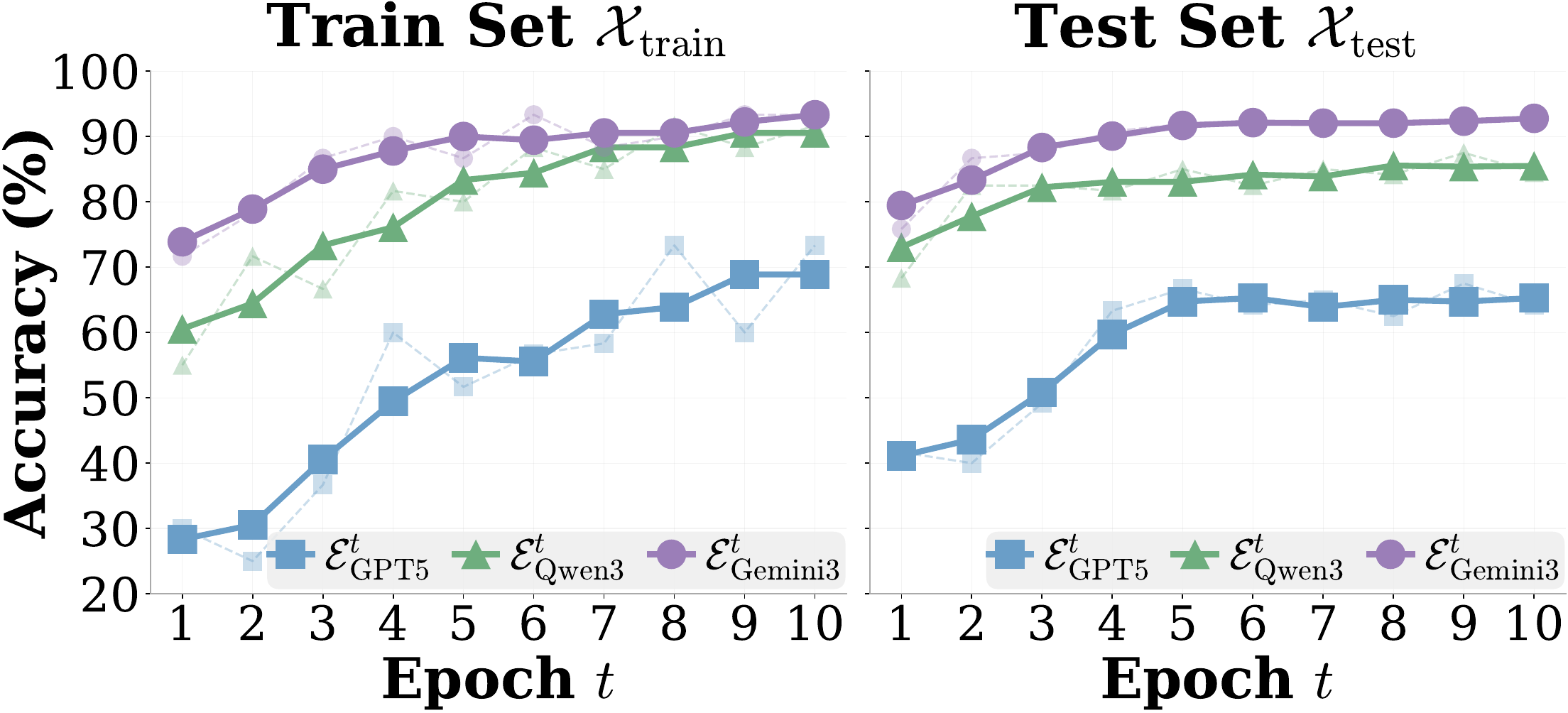}
    \caption{\textbf{Evolution dynamics on WorkBench.} Accuracy of $\mathcal{E}_t$ on $\mathcal{X}_{\mathrm{train}}$ (left) and $\mathcal{X}_{\mathrm{test}}$ (right) over 10 epochs. Dashed lines show mean over three trials while solid lines are moving averages to highlight trend.}
    \label{fig:evolution_dynamics}
\end{figure}

%% file: section/figure/ablation_OptiMAS.tex
\begin{figure}
    \centering
    \includegraphics[width=0.7\linewidth]{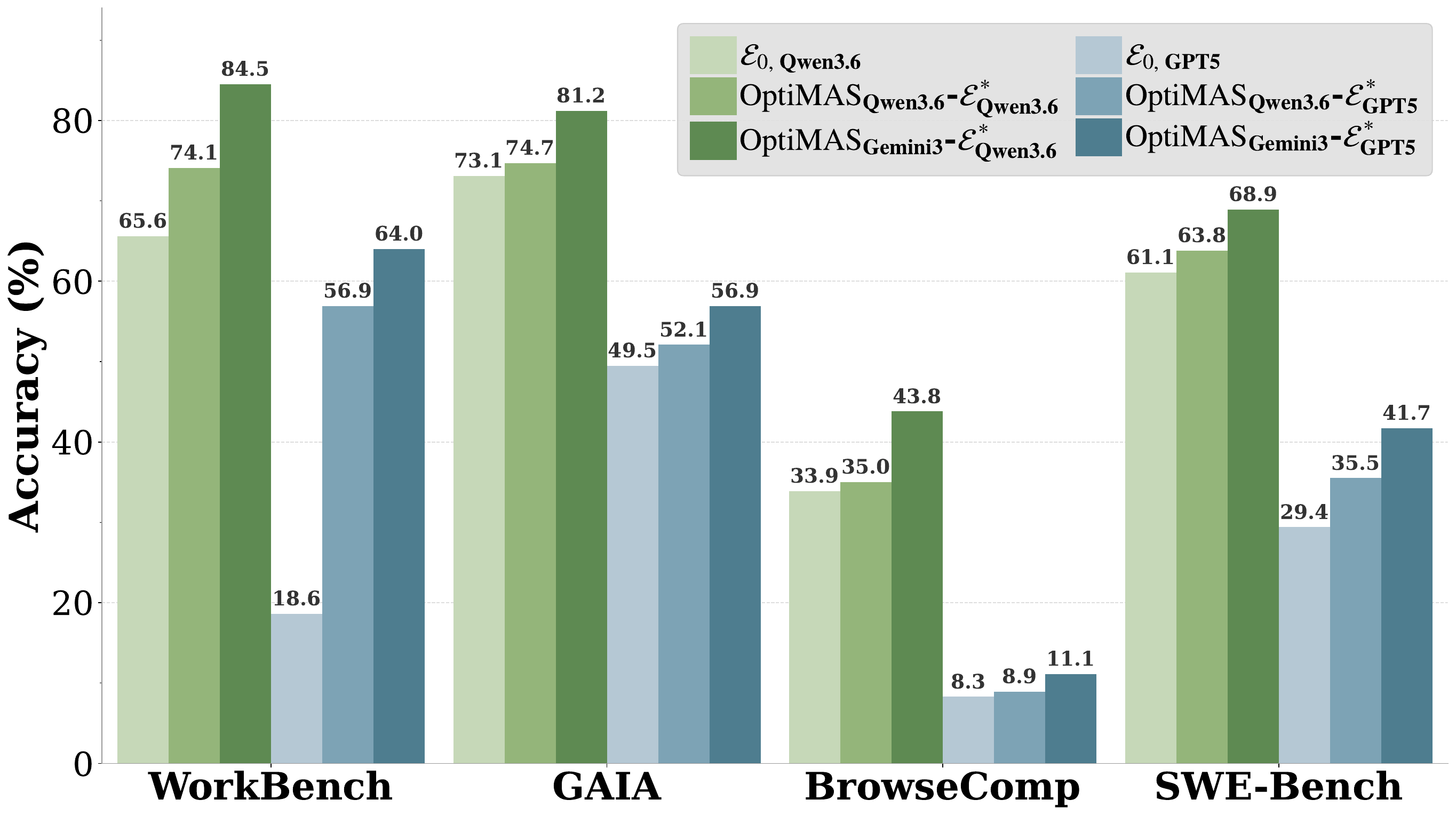}
    \caption{\textbf{Ablation analysis of optimizer backbones.} Performance is compared across combinations of two MAS backbones and two OptiMAS backbones.}
    \label{fig:ablation_optimas}
\end{figure}

%% file: section/table/ablations.tex
\begin{table}[t]
  \centering
  \caption{\textbf{Ablation analysis of framework components.} Bold indicates best results. ``\ding{53}'' signifies component removal, and ``BS'' represents batch size.}
  \label{tab:robustness_analysis}
  \scriptsize
  \setlength{\tabcolsep}{1.2pt}
  \renewcommand{\arraystretch}{1}
   \resizebox{0.5\linewidth}{!}{
  \begin{tabular}{l|cccc}
    \toprule
    \multirow{2}{*}{\centering \makecell{\textbf{Ablation} \\ \textbf{Settings}}}
      & \multicolumn{2}{c}{\textbf{WorkBench (\%) $\uparrow$}}
      & \multicolumn{2}{c}{\textbf{SWE-Bench (\%)$\uparrow$}} \\
    \cmidrule(lr){2-3}\cmidrule(lr){4-5}
      & GPT5 & Qwen3.6 & GPT5 & Qwen3.6  \\
    \midrule
    Initial $\mathcal{E}_0$ (baseline) & 18.6 & 65.6 & 29.4 & 61.1 \\
    \midrule
    \textbf{OptiMAS} w/ BS $n\!=\!10$ & \textbf{64.0\textsubscript{\textcolor{ForestGreen!80}{(+45.4)}}} & \textbf{84.5\textsubscript{\textcolor{ForestGreen!80}{(+18.9)}}} & \textbf{41.7\textsubscript{\textcolor{ForestGreen!80}{(+12.3)}}} & \textbf{68.9\textsubscript{\textcolor{ForestGreen!80}{(+7.8)}}} \\
    \ \ \ \ding{53} \ \ \ \ \ \ \ \ \ \ \  w/ BS $n\!=\!4$ & 55.0\textsubscript{\textcolor{ForestGreen!80}{(+36.4)}} & 81.1\textsubscript{\textcolor{ForestGreen!80}{(+15.5)}} &  24.4\textsubscript{\textcolor{red!80}{(-5.0)}} & 62.8\textsubscript{\textcolor{ForestGreen!80}{(+1.7)}} \\
    \ \ \ \ding{53} Validation Metric & 58.2\textsubscript{\textcolor{ForestGreen!80}{(+39.6)}} & 82.8\textsubscript{\textcolor{ForestGreen!80}{(+17.2)}} &  28.9\textsubscript{\textcolor{red!80}{(-0.5)}} & 64.4\textsubscript{\textcolor{ForestGreen!80}{(+3.3)}} \\
    \ \ \ \ding{53} Hypothesis Memory & 53.9\textsubscript{\textcolor{ForestGreen!80}{(+35.3)}} & 71.4\textsubscript{\textcolor{ForestGreen!80}{(+5.8)}} &  22.2\textsubscript{\textcolor{red!80}{(-7.2)}} &  53.3\textsubscript{\textcolor{red!80}{(-7.8)}} \\
    \ \ \ \ding{53} Adaptive Sampling & 60.3\textsubscript{\textcolor{ForestGreen!80}{(+41.7)}} & 75.1\textsubscript{\textcolor{ForestGreen!80}{(+9.5)}} &  24.4\textsubscript{\textcolor{red!80}{(-5.0)}} &  60.6\textsubscript{\textcolor{red!80}{(-0.5)}} \\
    \bottomrule
  \end{tabular}
  }
\end{table}

%% file: section/5_conclusion.tex
\section{Conclusions}
\label{sec:concl}

We presented OptiMAS, an optimization-based framework that achieves steady, continuous multi-agent system evolution on real-world, long-horizon agentic tasks. Our framework rests on three interconnected contributions. First, we established a unified ReAct-based infrastructure with a proactive assign-deliver protocol that reconciles a broad optimization scope with operational stability, enabling joint evolution of prompts, skills, orchestration, and tool configurations. Second, we reformulated MAS evolution as a continuous, data-driven optimization process with adaptive sampling and textual gradient analysis under strict train-validation-test separation to ensure generalization, replacing discrete branch-and-discard search with evidence-grounded progressive refinement. Third, we introduced the dual-track memory mechanism, comprising plan-oriented short-term memory for intra-step consistency and hypothesis-driven long-term memory for cross-step continuity, which transforms MAS evolution from undirected exploration into a principled, self-correcting process.
 
Extensive evaluation across four heterogeneous benchmarks and three backbone LLMs of varying scale and accessibility demonstrated that OptiMAS consistently improves over its initialization and achieves superior performance relative to both domain-specialized hand-crafted systems and existing evolutionary methods from a single task-agnostic configuration. We hope this work contributes to advancing the automated synthesis of deployable multi-agent systems.

%% file: section/7_limitations.tex
\section*{Limitations}
\label{sec:limitations}
 
\paragraph{Backbone capability requirements.}
The generality of our infrastructure, encompassing autonomous tool invocation, hierarchical sub-agent delegation, and proactive context management, places non-trivial demands on the backbone LLM's instruction-following and tool-calling proficiency. As a result, substantially weaker models may struggle to operate effectively within this framework. Extending compatibility to lighter-weight or edge-computation friendly models through simplified execution protocols or progressive capability scaffolding remains a valuable direction for future work.
 
\paragraph{Optimization exploration breadth.}
While OptiMAS yields consistent improvements across challenging agentic benchmarks, the current work represents an initial exploration of a vast design space. Drawing an analogy to the maturation of deep learning methodology, numerous dimensions remain to be investigated, including training curricula and scheduling strategies, dataset curation policies, multi-agent diversity and composition, and the inclusion of infrastructure components such as prompt templates and tools creation as first-class optimization targets. Each of these directions constitutes a promising avenue for extending the present framework.

%% file: section/8_appendix.tex
\newpage
\appendix

\input{section/appendix/A_benchmarks_curation}
\input{section/appendix/B_comparison}
\input{section/appendix/C_mas_config}
\input{section/appendix/D_experiment_hyperparameters}

\input{section/appendix/E_Infrastructure}
\input{section/appendix/F_OptiMAS_design}
\input{section/appendix/G_case_study}

%% file: section/appendix/A_benchmarks_curation.tex
\section{Benchmark $\mathcal{X}$ Curation and Partition}
\label{sec:appendix_benchmark_curation}

\subsection{Benchmarks $\mathcal{X}$ Curation}
Due to prohibitive computational resource requirement on long-horizon SWE-Bench-Verified and tremendous search and crawl required BrowseComp benchmarks, we construct fixed-size subsets, each compressing 100 instances, from the two established benchmarks by fixed-seed ramdon sampling to ensure reproducibility.

\paragraph{SWE-Bench-Verified.}
SWE-bench Verified~\cite{jimenez2024swe} is a human verified subset of the SWE-bench benchmark comprising real-world GitHub issues paired with ground-truth patches. We first obtain the full SWE-bench Verified dataset (totoally 500 instances) and sample a subset of $N=100$ instances using Python's \texttt{random.Random} with seed $s=42$. The resulting subset is stored as a static JSONL file and remains fixed throughout all experiments.

\paragraph{BrowseComp.}
BrowseComp~\cite{wei2025browsecomp} is a benchmark of challenging factual questions that require multi-step web browsing and cross-source verification, which totally compresses 1,266 queries. We also following the same random sampling rules with same random seed $s=42$ in SWE-Bench to sample $N=100$ queries from the BrowseComp validation set.

\paragraph{GAIA-text and WorkBench.} GAIA-text~\cite{mialon2023gaia} comprises 103 complex reasoning questions requiring mathematical calculation, file parsing, and cross-tool orchestration. WorkBench~\cite{styles2024workbench} contains 690 tasks spanning six workplace domains that test multi-step planning, tool manipulation, and hierarchical execution. As both benchmarks impose substantially lower computational overhead than SWE-Bench-Verified and BrowseComp, we retain their complete established sets as $\mathcal{X}$ without subsampling.

\subsection{Deterministic $\mathcal{X}_\text{train/val/test}$ Partition}
\label{sec:data_split_procedure}

\paragraph{SWE-Bench-Verified, BrowseComp, and GAIA.}
As these three benchmarks share comparable scales (100, 100, and 103 instances respectively), we apply a unified deterministic partitioning procedure to each:
\begin{enumerate}[leftmargin=*]
    \item[-] \textbf{Index generation.} Construct an index array $I = [0, 1, \ldots, N{-}1]$.
    \item[-] \textbf{Deterministic shuffle.} Shuffle $I$ in-place using Python's \texttt{random.Random} initialized with seed $s=42$.
    \item[-] \textbf{Contiguous partitioning.} Split the shuffled array into three contiguous segments with ratios $(r_{\mathrm{train}}, r_{\mathrm{val}}, r_{\mathrm{test}}) = (0.36, 0.04, 0.60)$:
    \begin{align}
        n_{\mathrm{train}} &= \lfloor N \cdot r_{\mathrm{train}} \rfloor,\\ 
        n_{\mathrm{val}} &= \lfloor N \cdot r_{\mathrm{val}} \rfloor, \\
        n_{\mathrm{test}} &= N - n_{\mathrm{train}} - n_{\mathrm{val}}.
    \end{align}
    \item[-] \textbf{Assignment.}
    \begin{align}
        \mathcal{X}_{\mathrm{train}} &= \{ x_{I[i]} \mid 0 \le i < n_{\mathrm{train}} \}, \notag \\
        \mathcal{X}_{\mathrm{val}} &= \{ x_{I[i]} \mid n_{\mathrm{train}} \le i < n_{\mathrm{train}} + n_{\mathrm{val}} \}, \notag \\
        \mathcal{X}_{\mathrm{test}} &= \{ x_{I[i]} \mid n_{\mathrm{train}} + n_{\mathrm{val}} \le i < N \}. \notag
    \end{align}
\end{enumerate}
Table~\ref{tab:data_split} summarizes the resulting partition sizes.
 
\begin{table}[h]
\centering
\caption{Dataset partitioning for evolutionary experiments.}
\label{tab:data_split}
\small
\resizebox{0.5\linewidth}{!}{
\begin{tabular}{lccccc}
\toprule
\textbf{Benchmark} & \textbf{Source size} & $|\mathcal{X}_{\mathrm{train}}|$ & $|\mathcal{X}_{\mathrm{val}}|$ & $|\mathcal{X}_{\mathrm{test}}|$ & \textbf{Total} $|\mathcal{X}|$ \\
\midrule
SWE-Bench-Verified & 500 & 36 & 4 & 60 & 100 \\
BrowseComp & 1,266 & 36 & 4 & 60 & 100 \\
GAIA-text & 103 & 37 & 4 & 62 & 103 \\
\bottomrule
\end{tabular}
}
\end{table}

\subsection{WorkBench}
WorkBench~\cite{styles2024workbench} comprises 690 task instances spanning six functional domains: \textit{Analytics} (120), \textit{Calendar} (110), \textit{CRM} (80), \textit{Email} (90), \textit{Multi-Domain} (210), and \textit{Project Management} (80). Unlike the three benchmarks above, WorkBench exhibits significant domain imbalance, so we apply stratified sampling to ensure uniform domain coverage in the training and validation splits. For each domain, instances are shuffled with seed $s=42$, and the first 10 are assigned to $\mathcal{X}_\text{train}$, the next 2 to $\mathcal{X}_\text{val}$, with all remaining instances forming $\mathcal{X}_\text{test}$. This yields 60 training, 12 validation, and 618 test instances. Table~\ref{tab:data_splits} summarizes the per-domain partition.
 
\begin{table}[t]
\centering
\caption{WorkBench dataset partition via stratified sampling (seed $s=42$).}
\label{tab:data_splits}
\resizebox{0.5\linewidth}{!}{
\begin{tabular}{lrrrr}
\toprule
\textbf{Domain} & \textbf{Source} & $|\mathcal{X}_\text{train}|$ & $|\mathcal{X}_\text{val}|$  & $|\mathcal{X}_\text{test}|$  \\
\midrule
Analytics               & 120 & 10 & 2 & 108 \\
Calendar                & 110 & 10 & 2 &  98 \\
CRM                     &  80 & 10 & 2 &  68 \\
Email                   &  90 & 10 & 2 &  78 \\
Multi-Domain            & 210 & 10 & 2 & 198 \\
Project Management      &  80 & 10 & 2 &  68 \\
\midrule
\textbf{Total}          & \textbf{690} & \textbf{60} & \textbf{12} & \textbf{618} \\
\bottomrule
\end{tabular}
}
\end{table}

\subsection{Role of Each Split in Evolutionary Training}
 
\paragraph{Training split $\mathcal{X}_{\mathrm{train}}$.}
At each optimization step, the adaptive sampler draws a batch of $n$ instances from $\mathcal{X}_{\mathrm{train}}$ via probability-weighted sampling, prioritizing queries that failed in the previous step while filling remaining slots from the full training pool. The resulting trajectories and evaluation outcomes constitute the forward-pass output and are routed to the optimizer for backward-pass analysis. The sampled subset varies across steps, but the candidate pool remains fixed.
 
\paragraph{Validation split $\mathcal{X}_{\mathrm{val}}$.}
The identical, small-scale $\mathcal{X}_{\mathrm{val}}$ is evaluated at every step using the current MAS configuration, providing a consistent overfitting signal. The optimizer receives the validation accuracy and its step-over-step delta as backward-pass input but has no access to validation trajectories, preventing it from fitting to validation instances.
 
\paragraph{Test split $\mathcal{X}_{\mathrm{test}}$.}
Held out entirely during evolution. Neither the optimizer nor any training component observes test instances or outcomes. The test set is used exclusively for final evaluation of the evolved MAS.
 
\subsection{Reproducibility Notes}
\begin{itemize}[leftmargin=*]
    \item A fixed random seed $s=42$ governs both subset sampling and train/val/test partitioning, ensuring deterministic reproduction given the same source dataset. All curated data and code will be publicly released.
    \item Partitioning is performed once before training and remains fixed across all epochs, steps, and benchmarks.
    \item Split ratios and seed are configurable via command-line arguments (e.g., \texttt{-{}-split-ratios 0.36 0.04 0.60} and \texttt{-{}-seed 42}).
    \item Python's \texttt{random.Random(42).shuffle()} is the sole source of randomness in partitioning, with no external library dependencies, ensuring cross-platform determinism.
\end{itemize}

%% file: section/appendix/B_comparison.tex
\section{Baseline Details}
\label{sec:appendix_baselines}

We evaluate OptiMAS against baselines spanning two paradigms: hand-crafted multi-agent systems designed by human engineers, and automated frameworks that algorithmically discover or optimize agent architectures.

\subsection{Hand-crafted Multi-Agent Systems}

\begin{itemize}[left=0pt]
    \item \textbf{SMoA}~\cite{li2025smoa}: A general-purpose Sparsified Mixture of Agents framework that harnesses collaborative capabilities of multiple heterogeneous language models. To reduce the computational overhead typical of dense multi-agent configurations, SMoA incorporates dynamic response selection and early stopping mechanisms to sparsify information flows, balancing multi-agent synergy with execution efficiency.
    
    \item \textbf{Multi-Agent Debate}~\cite{du2023improving}: A general-purpose cooperative framework in which multiple independent language model agents address a problem simultaneously and iteratively refine their reasoning through successive rounds of structured peer critique. This process enables cross-verification of facts and systematic reduction of individual agent hallucinations.
    
    \item \textbf{SWE-Agent}~\cite{yang2024sweagent}: An open-source software engineering agent framework designed to resolve real-world code defects within complex repositories. It introduces a custom Agent-Computer Interface (ACI) that provides optimized terminal interaction, repository navigation, and file-editing tools, simplifying the action space for LLM-driven code repair. In our experiments, we re-implement SWE-Agent atop our unified infrastructure while retaining its official prompts and tool configurations, ensuring both ease of integration and fair comparison. We use SWE-Agent as a domain-specialized baseline for \emph{SWE-Bench-Verified}.

     \item \textbf{Tongyi-DeepResearch}~\cite{tongyidr}: An open-source agentic framework with a task-specific finetuned LLM backbone (30B total, 3.3B activated) developed by Tongyi Lab for long-horizon, deep information-seeking tasks. It employs an end-to-end agentic training paradigm that unifies continual pre-training on large-scale agentic interaction data with multi-turn reinforcement learning via a customized GRPO framework. At inference, it supports both a standard ReAct mode and an IterResearch-based test-time scaling mode. We use Tongyi-DR framework with general-purposed LLM backbone as a domain-specialized baseline for \emph{GAIA-text} and \emph{BrowseComp}.
\end{itemize}

\subsection{Automated Agent Evolution Methods}

\begin{itemize}[left=0pt]
    \item \textbf{ADAS}~\cite{hu2024automated}: The Automated Design of Agentic Systems framework formalizes open-ended discovery of agent architectures in a code-level search space. A meta-agent iteratively proposes novel programmatic agentic components, which are then evaluated in isolated sandbox environments, progressively building an archive of discovered designs.
    
    \item \textbf{EvoAgent}~\cite{yuan2025evoagent}: An automated framework that extends single-agent systems into multi-agent collaborative networks through evolutionary operators including semantic mutation, crossover, and quality-based selection. EvoAgent generates a diverse population of specialized agent personas from a single user-provided template, removing the need for manual orchestration design.
    
    \item \textbf{DGM}~\cite{zhang2025darwin}: The Darwin G\"{o}del Machine is a self-evolving system that unifies functional execution logic and self-modification routines into editable programs. Through iterative cycles of programmatic mutation, sandboxed evaluation, and archive-based exploration, DGM enables agents to continuously refine their own code and capabilities beyond their initial architectural boundaries.
\end{itemize}

%% file: section/appendix/C_mas_config.tex
\section{$\mathcal{E}_0$ Configurations}
\label{sec:appendix_mas_config}

This section documents the initial (epoch-zero) Multi-Agent System configurations used in our experiments on the all four benchmarks. Both configurations adopt a \emph{single-agent} architecture $\mathcal{E}_0$ as the starting point for evolutionary optimization, comprising one general agent (\texttt{worker}) with no sub-agents. The evolutionary process may subsequently introduce additional agents, create new skills, modify prompts, and adjust tool permissions as the OptiMAS identifies structural or behavioral improvements. The initial configurations therefore represent the \emph{minimal viable MAS} from which the optimizer begins its search.

\subsection{General $\mathcal{E}_0$ Configuration}

\subsubsection{Topology and Iteration Budget}
The MAS consists of a single entry agent (\texttt{worker}) with no sub-agents. The iteration budget is set to 40 reasoning steps per phase and 100 total steps across all phases, providing sufficient room for multi-phase exploration, patch construction, and verification.

\subsubsection{LLM/VLM Backbone}
The primary reasoning backbones span from GPT-5-Nano (128K), Qwen-3.6-35B-A3B(262K), Gemin-3-Flash(1M) under different experimental settings. The same model is used for internal AI tools (e.g., \texttt{text\_qa} for document understanding and context compression). The vision-language model for \texttt{image\_qa} is GLM-4.6V-Flash under Qwen-3.6-35B-A3B setting, and GPT-4o-mini under GPT-5-Nano and Gemin-3-Flash settings.

\subsubsection{Initial System Prompt}
The initial system prompt establishes the agent's role and environment:

\begin{promptbox}
You are an autonomous agent that solves tasks by using tools.

Today's date: \{today\_date\}

\#\# System protocols (mandatory)

The following protocols are fully loaded below. You must follow them.

\{system\_protocols\}

\#\# Optional skills

You may load additional skills with the `load\_skill` tool if needed. Only each skill's name and description appear below; the full instruction body is returned in the tool response when you call `load\_skill`.

\{optional\_skills\}

\end{promptbox}

The value in `\{\}' placeholders will be resolved upon MAS instantiation. The initial system prompt is intentionally minimal. It defines the agent's role and execution environment but does not prescribe any specific workflow, debugging strategy, or patch construction methodology. This design leaves the behavioral search space maximally open for the optimizer to populate through skill creation and prompt refinement. The system prompt also contains two template regions, \texttt{\$\{system\_protocols\}} and \texttt{\$\{optional\_skills\}}, into which the framework injects the system skill bodies and optional skill descriptions at runtime.

\subsubsection{Initial Message Template}
\begin{promptbox}
\{task\}
\end{promptbox}
The initial message template consists solely of the \texttt{\$\{task\}} placeholder, which is replaced at runtime with the specific queries description directly given by benchmark. No additional scaffolding or structured instructions are provided, leaving the agent to rely entirely on its system prompt and skills for guidance.

\subsubsection{System Protocols}
Based on the execution mechanism and strong coupled work control tools of our infrastructure, the foundational system protocol is inlined in default into the system prompt to ensure the Agent knows how to work and deliver solution upon completion.

\begin{promptbox}

name: phase-transition

description: Decide when to call phase\_done\_and\_summarize versus task\_done.

    - Call `phase\_done\_and\_summarize' when there is remaining work or the context is getting long. This compresses history and continues execution.

    - Call `task\_done' only when the task is fully complete.
\end{promptbox}

\subsubsection{Available Tools}

The general-purpose $\mathcal{E}_0$ is equipped with 10 tools:
\begin{itemize}[leftmargin=*]
    \item \textbf{File operations}: \texttt{read\_file}, \texttt{write\_file}, \texttt{file\_edit}, \texttt{grep\_file\_content}, \texttt{list\_file\_tree} for navigating, reading, and modifying artifacts in its own workspace.
    \item \textbf{Content analysis}: \texttt{text\_qa} for LLM-powered question answering over long files with automatic sliding-window chunking.
    \item \textbf{Workflow control}: \texttt{load\_skill}, \texttt{task\_done}, \texttt{phase\_done\_and\_summarize} for optional skill loading, task completion, and phase transitions.
\end{itemize}

\subsubsection{Optional Skills}
The initial configuration includes no optional skills. The OptiMAS may create and register optional skills during evolution to address recurring failure patterns.

\subsection{Specific $\mathcal{E}_0$ Configurations}
Based on the general-purpose configurations of $\mathcal{E}_0$, some specific tools are required to ensure the corresponding task can be completed, otherwise the worse initial $\mathcal{E}_0$ performance could be necessary tools absence.

\subsubsection{SWE-Bench $\mathcal{E}_0$}
Specific tools for $\mathcal{E}_0$ on SWE-Bench-Verified benchmarking:
\begin{itemize}[leftmargin=*]
\item \textbf{Shell execution} (1 tool): \texttt{shell\_exec} for running commands (e.g., \texttt{pytest}, \texttt{git diff}, \texttt{python}) inside a Docker container with the repository's environment pre-configured.
\end{itemize}

\subsubsection{BrowseComp $\mathcal{E}_0$}
Specific tools for $\mathcal{E}_0$ on BrowseComp benchmarking:
\begin{itemize}[leftmargin=*]
    \item \textbf{Web and network} (3 tools): \texttt{search} (web search via Serper API), \texttt{crawl} (page content retrieval via Jina Reader), \texttt{download} (file downloading to workspace), enabling multi-step web research, source retrieval, and cross-referencing.
\end{itemize}

\subsubsection{GAIA-text $\mathcal{E}_0$}
Specific tools for $\mathcal{E}_0$ on GAIA-text benchmarking:
\begin{itemize}[leftmargin=*]
    \item \textbf{Web and network} (3 tools): \texttt{search} (web search via Serper API), \texttt{crawl} (page content retrieval via Jina Reader), \texttt{download} (file downloading to workspace), enabling multi-step web research, source retrieval, and cross-referencing.
\end{itemize}

\subsubsection{WorkBench $\mathcal{E}_0$}
The initial environment $\mathcal{E}_0$ for WorkBench equips the agent with 29 tools spanning six functional categories, mirroring the complete API surface of the WorkBench sandbox~\cite{styles2024workbench}:
\begin{itemize}[leftmargin=*]
    \item \textbf{Calendar management} (5 tools): \texttt{create\_event}, \texttt{delete\_event}, \texttt{update\_event}, \texttt{get\_event\_information\_by\_id}, and \texttt{search\_events}. These provide full CRUD access over a calendar database, enabling the agent to schedule, modify, query, and remove events by ID, name, participant, or time range.
    \item \textbf{Email operations} (6 tools): \texttt{send\_email}, \texttt{delete\_email}, \texttt{forward\_email}, \texttt{reply\_email}, \texttt{get\_email\_information\_by\_id}, and \texttt{search\_emails}. Beyond basic send and delete, the inclusion of \texttt{forward} and \texttt{reply} as first-class operations reflects WorkBench tasks that require multi-hop communication chains (e.g., forwarding a searched email to a looked-up contact).
    \item \textbf{Web analytics} (6 tools): \texttt{create\_plot}, \texttt{total\_visits\_count}, \texttt{engaged\_users\_count}, \texttt{traffic\_source\_count}, \texttt{get\_average\_session\_duration}, and \texttt{get\_visitor\_information\_by\_id}. This toolkit combines aggregation queries (visits, engagement, traffic sources, session duration over date ranges) with a visualization tool, covering the analytical reasoning tasks in the benchmark.
    \item \textbf{Project management} (5 tools): \texttt{create\_task}, \texttt{delete\_task}, \texttt{update\_task}, \texttt{get\_task\_information\_by\_id}, and \texttt{search\_tasks}. Tasks are organized by boards (\textit{Back end}, \textit{Front end}, \textit{Design}), lists (\textit{Backlog}, \textit{In Progress}, \textit{In Review}, \textit{Completed}), and assignees, requiring the agent to navigate structured project hierarchies.
    \item \textbf{Customer relationship management} (4 tools): \texttt{add\_customer}, \texttt{update\_customer}, \texttt{delete\_customer}, and \texttt{search\_customers}. The CRM toolkit supports multi-field filtering (name, email, product interest, status, contact dates) and enforces domain constraints on status and product categories, testing the agent's ability to respect enumerated value restrictions.
    \item \textbf{Company directory} (1 tool): \texttt{find\_email\_address} resolves employee names to email addresses. This cross-cutting utility is essential for multi-domain tasks where the agent must first look up a contact before performing calendar, email, or CRM operations---a common pattern in WorkBench's \textit{multi\_domain} category.
    \item \textbf{Agent control} (2 tools): \texttt{task\_done} signals task completion, and \texttt{phase\_done\_and\_summarize} enables phased execution with intermediate summaries. These are system-level primitives for the MAS coordination protocol rather than domain-specific APIs.
\end{itemize}

\paragraph{Minimalist Initialization Policy.}
The initialization of the agentic toolkits is intentionally governed by a \emph{minimalist design principle}. Rather than seeding the ecosystem with domain-specific heuristics, specialized sequential workflows, or optional high-level skills, we equip each agent exclusively with generic foundational primitives essential for baseline operations (i.e., fundamental file manipulation, context routing, and standardized execution termination). This strict structural constraint maximizes the behavioral exploration space accessible to the evolutionary optimizer. Consequently, it guarantees that any observed trajectory optimizations and performance increments during the evolutionary training phase are strictly attributable to the optimizer’s structural interventions rather than pre-engineered initial configurations.

\paragraph{Orthogonal Tool-Isolation and Data Leakage Prevention.}
To preserve evaluation integrity and ensure strict alignment with the underlying benchmarks, we enforce an orthogonal tool-isolation protocol across distinct task modalities. Specifically, we prohibit web search capabilities for the multi-agent systems (MAS) evaluated on \textbf{SWE-Bench-Verified}, while symmetrically blocking shell execution and programmatic code-generation tools for the MAS tested on \textbf{BrowseComp} and \textbf{GAIA-text}. 

The structural necessity of this constraint stems from a telemetry vulnerability discovered during early-stage exploration. When unconstrained, the search-enabled MAS could heuristically locate the remote ground-truth data repositories via open-web queries. It would then attempt to bypass real-time reasoning by downloading and programmatically parsing the dataset files (e.g., extracting serialized labels from cloud-hosted tables via synthesized Python scripts). By restricting shell execution and coding toolkits in web-centric benchmarks, the agents are structurally blocked from processing production-grade data formats such as Parquet (frequently deployed on Hugging Face repositories), as our basic file-reading utilities lack native handlers for compressed binary schemas. This restriction forces the evolved MAS to perform authentic, zero-leakage information foraging and verification. Symmetrically, stripping web-search primitives from the code-centric \textbf{SWE-Bench-Verified} pipeline compels the agents to rely entirely on static repository comprehension and local unit-test validation, preventing any empirical contamination from external web-hosted patches.

%% file: section/appendix/D_experiment_hyperparameters.tex
\section{Experiment Hyperparameters}
\label{sec:appendix_experiment_implementation}

This section details the numerical hyperparameters governing our evolutionary optimization framework. We categorize these parameters into three dimensional layers: \emph{training loop} configurations (defining the temporal optimization boundaries), \emph{adaptive sampling} coefficients (modulating the density-weighted batch construction). All parameters are maintained uniformly across all benchmark evaluations unless explicitly stated otherwise.

\subsection{Training Horizon and Optimization Boundaries}
\label{sec:hp_training_loop}

To accommodate variations in sequence length and task density across the evaluated benchmarks, we adopt the maximum number of \emph{epochs} rather than a fixed step budget as the optimization termination criterion. This design choice prevents empirical bias stemming from heterogeneous dataset scales while holding the candidate batch size $n$ constant. For a designated training split $\mathcal{X}_\text{train}$, the step cardinality per epoch is formalized as follows:
\begin{equation}
    \text{step} = \left\lceil \frac{|\mathcal{X}_\text{train}|}{n} \right\rceil.
\end{equation}
Given the heavy computational requirements of long-horizon evolution, the optimization horizon is tightly bounded at exactly 10 epochs across all four benchmarks, guaranteeing that the evolving MAS exhaustively explores the complete support dataset $\mathcal{X}_\text{train}$ ten times. To preserve structural equity during comparison, all search-based and evolutionary baseline counterparts (e.g., DGM~\cite{zhang2025darwin}, which mandates an offspring generation policy of two candidate MAS structures per cycle) are configured to execute for precisely ten generations, with all remaining parameters adhering strictly to their officially calibrated settings.

\subsection{Adaptive Sampling Parameters and Design Rationales}
\label{sec:hp_adaptive_sampling}

As formulated in Algorithm~\ref{alg:adaptive_sampling}, the adaptive sampling subsystem introduces a localized, priority-weighted batch construction mechanism. At each gradientless optimization step, a designated fraction $\phi$ of the batch capacity is reserved exclusively to re-execute queries that failed during the immediate historical iteration. The remaining allocation slots are dynamically populated via probability-weighted sampling across the entire training split $\mathcal{X}_\text{train}$. Sampling weight updates are parameterically controlled by multiplicative scaling factors $\alpha$ and $\beta$, bounded within the closed interval $[p_{\min}, p_{\max}]$ to mitigate long-term distributional drift. To prevent catastrophic weight saturation, a periodic hard reset interval $\tau$ is enforced. The specific parameter calibrations are consolidated in Table~\ref{tab:hp_sampler}.

\begin{table*}[h]
\centering
\caption{Parameter configurations for the adaptive sampling subsystem.}
\label{tab:hp_sampler}
\small
\begin{tabular}{lll}
\toprule
\textbf{Hyperparameter} & \textbf{Value} & \textbf{Mathematical Functional Description} \\
\midrule
$\phi$ (Retry Ratio) & 0.5 & Mass allocation ratio reserved for historic failed queries \\
$\alpha$ (Decay Factor) & 0.6 & Multiplicative discount multiplier for successful queries \\
$\beta$ (Boost Factor) & 1.2 & Multiplicative scaling multiplier for failed queries \\
$p_{\min}$ (Weight Floor) & 0.2 & Lower-bound boundary for sample probability clamping \\
$p_{\max}$ (Weight Ceiling) & 2.0 & Upper-bound boundary for sample probability clamping \\
$\tau$ (Reset Interval) & 2 epochs & Temporal stride between global uniform weight resets \\
$w_0$ (Initial Weight) & 1.0 & Uniform initial scalar assigned to all dataset instances \\
\bottomrule
\end{tabular}
\end{table*}

The calibration of these hyperparameter values is dictated by structural design rationales aimed at stabilizing trajectory exploration. The clamping boundaries ($p_{\min} = 0.2, p_{\max} = 2.0$) restrict the maximum probability discrepancy between consistently erroneous tasks and successfully resolved tasks to a factor of $\times 10$, ensuring the sampling distribution remains non-degenerate and safe from saturation. Correspondingly, the step-wise modifiers $\alpha = 0.6$ and $\beta = 1.2$ are mathematically mirrored to ensure that the lower probability boundary is systematically reached after five consecutive successes, while the upper ceiling is attained after five consecutive execution failures. 

Crucially, as established by the global framework ablations in Section~\ref{sec:exper}, the macroscopic trajectory improvements are robust to marginal variations in these auxiliary local coefficients. The validation of the global adaptive mechanism, coupled with the hypothesis-driven long-term memory module, fully demonstrates structural stability without requiring hyperparameter grid-searches.

\subsection{Hyperparameter Synthesis}

Table~\ref{tab:hp_all} summarizes the key global hyperparameter configurations implemented across our entire framework infrastructure.

\begin{table*}[t]
\centering
\caption{Consolidated global hyperparameter specifications.}
\label{tab:hp_all}
\small
\begin{tabular}{lll}
\toprule
\textbf{Operational Category} & \textbf{Hyperparameter Layer} & \textbf{Calibrated Value} \\
\midrule
\multirow{2}{*}{Training Loop Boundaries}
    & Max Optimization Horizon $E$ & 10 epochs \\
    & Batch Capacity $B$ & 10 \\
\midrule
\multirow{6}{*}{Adaptive Sampling Subsystem}
    & Retry Ratio $\phi$ & 0.5 \\
    & Decay Factor $\alpha$ & 0.6 \\
    & Boost Factor $\beta$ & 1.2 \\
    & Weight Floor $p_{\min}$ & 0.2 \\
    & Weight Ceiling $p_{\max}$ & 2.0 \\
    & Global Reset Interval $\tau$ & $2 \times \lceil |\mathcal{X}_{\text{train}}| / n \rceil$ steps \\
\bottomrule
\end{tabular}
\end{table*}

%% file: section/appendix/E_Infrastructure.tex
\section{Infrastructure}
\label{sec:appendix_infrastructure}

This section describes the infrastructure that supports the our configuration-based agent instantiate and agent's interaction with the environment.

Our framework adopts an industry-standard client--server architecture grounded in the \emph{Model Context Protocol} (MCP), providing a modular, extensible, and deployment-ready foundation for autonomous software engineering agents.

\subsection{Infrastructure Overview}
\label{sec:infra_overview}

The system is organized as a decoupled two-tier architecture comprising an \textbf{Agent Client} and a extensible \textbf{Tools Server}, communicating via the Model Context Protocol over Streamable HTTP transport. This separation confers several advantages:

\begin{itemize}[leftmargin=*]
    \item \textbf{Deployment Flexibility.} The agent client (hosting the LLM reasoning loop) and the tool server (hosting environment interactions) can be deployed on the same machine or distributed across heterogeneous infrastructure. This enables scenarios such as running the agent on GPU-equipped nodes while the tool server operates on standard compute instances co-located with development environments.
    \item \textbf{Multi-Tenancy and Isolation.} The server manages concurrent sessions with per-session workspace isolation, supporting parallel agent executions with no cross-contamination. Each session maintains its own file system scope, process state, and lifecycle management.
    \item \textbf{Protocol Compliance.} By building on MCP, our infrastructure ensures interoperability with the broader ecosystem of MCP-compatible tools and clients, facilitating integration with third-party services and framework.
    \item \textbf{Security.} The server enforces API key authentication, session validation (via HMAC-based session identifiers), and workspace sandboxing, ensuring that agent actions are confined to authorized scopes.
\end{itemize}

The agent client implements a ReAct-style reasoning loop that generates tool calls dispatched to the MCP server. Tool schemas are dynamically discovered at session initialization via the protocol's \texttt{tools\_accessible} capability, enabling the agent to adapt to varying server configurations without client-side modification.

\subsection{Tool Server Capabilities}
\label{sec:tool_capabilities}

The tool server provides a comprehensive suite of operations organized into five functional categories, collectively enabling the agent to perform long horizon, multi-step agentic tasks autonomously.

\paragraph{File Operations.}
The file operations module provides fine-grained control over the agent's workspace file system:
\begin{itemize}[leftmargin=*]
    \item \emph{Reading}: full-file and line-range reading with line number annotations, chunked reading for large-file question answering, and image reading with base64 encoding for vision-language model integration.
    \item \emph{Writing}: full-file creation and overwriting, append mode, and line-level insertion for surgical modifications.
    \item \emph{Editing}: targeted insertion, replacement, and deletion operations with line-level precision, as well as block-level replacement for refactoring workflows.
    \item \emph{Search and Navigation}: content-level regular expression search across the workspace, recursive file tree listing with configurable depth, and file move/rename/delete operations.
    \item \emph{Rich Format Support}: transparent extraction of textual content from structured document formats (PDF, DOCX, PPTX, XLSX, CSV, HTML) via an integrated document reader, enabling the agent to process heterogeneous file types uniformly.
\end{itemize}

\paragraph{System Execution.}
The shell execution module supports both \emph{local} and \emph{containerized} execution modes. In local mode, commands execute within the session workspace directory under permission identification. In containerized mode, which is used for reproducible evaluation on benchmarks such as SWE-bench, commands are transparently proxied into a Docker container with the appropriate environment pre-activated (e.g., conda environments with correct dependency versions). This dual-mode design allows the same agent logic to operate across development and evaluation contexts without modification.

\paragraph{Network Operations.}
The framework integrates two complementary information retrieval services:
\begin{itemize}[leftmargin=*]
    \item \emph{Web Search}: powered by the Serper API, supporting multiple search modalities (web, images, video, academic literature). A server-side token-bucket rate limiter ensures compliance with API rate constraints under concurrent multi-agent workloads.
    \item \emph{Web Crawling}: powered by the Jina Reader API, providing structured extraction of web page content in Markdown format. The crawler supports batch URL processing with automatic file persistence to the agent's workspace, and includes retry logic with configurable timeouts for robustness against transient network failures.
    \item \emph{File Download}: asynchronous streaming download of remote resources (including images, datasets, and documentation) with content validation, size limits, and automatic MIME type detection.
\end{itemize}

These capabilities enable the agent to gather external information, such as API documentation, library changelogs, issue discussions, and code examples, as part of its problem-solving workflow, significantly expanding the scope of tasks it can address autonomously.

\paragraph{Coding Operations.}
Dedicated coding tools provide language-aware support for the software engineering workflow:
\begin{itemize}[leftmargin=*]
    \item \emph{Syntax and Compilation Diagnostics}: static analysis of Python source code for syntax errors and compilation issues, returning structured feedback with line numbers, error types, and source excerpts.
    \item \emph{Function-Level Testing}: in-process execution of test functions against agent-written code with configurable timeouts, enabling rapid feedback cycles without full test suite overhead.
\end{itemize}

\subsection{Containerized Execution for Reproducible Evaluation}
\label{sec:docker_integration}

For benchmark evaluation, the framework provides seamless Docker container integration. Given a benchmark instance identifier (e.g., a SWE-bench problem ID), the system automatically:
\begin{enumerate}[leftmargin=*]
    \item Resolves and pulls the corresponding Docker image containing the exact repository snapshot and dependency environment.
    \item Extracts the working directory from the container image to the host file system for workspace initialization.
    \item Starts a new container with the workspace bind-mounted, enabling the agent to read and modify files via the file operations module while executing commands inside the container's pre-configured environment.
    \item Manages the container lifecycle (creation, monitoring, cleanup) to prevent resource leaks in batch evaluation scenarios.
\end{enumerate}

This approach ensures that the agent operates in an environment identical to the one used by human developers for the target repository, eliminating environment mismatch as a source of evaluation noise.

\subsection{Middleware Architecture}
\label{sec:middleware}

The server employs a layered middleware pipeline that intercepts all tool invocations, providing cross-cutting concerns without polluting individual tool implementations:

\begin{itemize}[leftmargin=*]
    \item \textbf{Authentication Middleware}: validates API keys on every request, preventing unauthorized access.
    \item \textbf{Session Middleware}: resolves and validates session identifiers, scopes tool invocations to the correct workspace, and maintains session liveness tracking for automatic expiration and cleanup.
    \item \textbf{Logging Middleware}: records structured invocation logs (tool name, arguments, success/failure, duration) in JSONL format, enabling fine-grained performance analysis and debugging of agent behavior.
\end{itemize}

The middleware chain executes in a fixed, deterministic order (authentication $\rightarrow$ session resolution $\rightarrow$ logging $\rightarrow$ tool execution), ensuring consistent behavior across all tool types.

\subsection{Scalability and Concurrency}
\label{sec:scalability}

The infrastructure is designed for concurrent multi-agent execution, which is essential for training workflows where multiple agent instances solve different problems in parallel:

\begin{itemize}[leftmargin=*]
    \item \textbf{Asynchronous I/O}: the server is built on an ASGI framework with full asynchronous support, enabling efficient handling of concurrent tool invocations without thread-per-request overhead.
    \item \textbf{Session Isolation}: each agent instance operates in a unique session with isolated workspace, process context, and container binding, preventing interference between concurrent executions.
    \item \textbf{LLM Concurrency Control}: a process-wide semaphore limits concurrent LLM API calls, preventing GPU or API rate limit saturation when multiple agents share a backend model deployment.
    \item \textbf{Rate-Limited External APIs}: outbound calls to external services (web search, crawling) are governed by token-bucket rate limiters, ensuring compliance with provider rate limits even under high-concurrency workloads.
\end{itemize}

\subsection{Generality and Extensibility}
\label{sec:extensibility}

The framework's modular design supports straightforward extension to new task domains and tool ecosystems:

\begin{itemize}[leftmargin=*]
    \item \textbf{Plugin-Style Tool Registration}: new tools are added by implementing a registration function and declaring parameter schemas via standard JSON Schema annotations. No modifications to the core framework or agent logic are required.
    \item \textbf{Multi-Domain Applicability}: the tool suite is not specific to any single benchmark or programming language. The same infrastructure supports software engineering tasks, mathematical reasoning, question answering, and web-based research, demonstrating the framework's generality.
    \item \textbf{Configuration-Driven Behavior}: agent capabilities, tool access, iteration budgets, and context limits are all governed by declarative YAML configuration, enabling rapid experimentation with different agent profiles without code changes.
    \item \textbf{Hierarchical Multi-Agent Support}: the architecture supports multi-agent delegation, where a parent agent can spawn sub-agents with independent LLM instances, tool sets, and conversation contexts, enabling sophisticated task decomposition strategies.
\end{itemize}

In summary, the infrastructure layer provides an industrial-grade, protocol-compliant foundation that abstracts the complexity of environment interaction behind a clean tool interface. This enables the agent's reasoning layer to focus on high-level problem solving while the infrastructure ensures reliable, secure, and scalable execution of its actions.

\subsection{Context Management Mechanism}
\label{sec:appendix_cmm}

A central challenge in deploying LLM-based agents for long-horizon agentic tasks, such as web searching and software engineering, is \emph{context window management}. The finite context length of the underlying language model imposes an inherent upper bound on the amount of information available during each reasoning step. As the agent interacts with tools, accumulates observations, and formulates plans, the conversation history grows monotonically, and naive truncation risks discarding critical state information. We address this challenge through a principled, multi-tier context lifecycle management framework that orchestrates \emph{proactive} (agent-initiated) and \emph{reactive} (system-enforced) phase transitions alongside \emph{semantic} and \emph{structural} compression mechanisms, ensuring that the agent retains the maximal amount of task-relevant information within its capacity constraints.

\subsubsection{Context Lifecycle Architecture}
\label{sec:context_lifecycle}

We model the evolution of the agent's conversational context as a managed lifecycle governed by a three-tier capacity envelope: a \textbf{soft limit} $C_{\text{soft}}$, a \textbf{hard limit} $C_{\text{hard}}$, and a \textbf{force limit} $C_{\text{force}}$ (with $C_{\text{soft}} < C_{\text{hard}} < C_{\text{force}}$). Each tier triggers progressively stronger interventions, forming a graceful degradation cascade rather than an abrupt cutoff.

Formally, let $\mathcal{H}_t = [m_1, m_2, \ldots, m_t]$ denote the conversation history at step $t$, and let $S(\mathcal{H}_t) = \sum_{i=1}^{t} |m_i|$ be the estimated payload size (measured in characters, with heuristic token approximation $\hat{T} \approx S/3 + 4|\mathcal{H}_t|$). At each iteration of the ReAct loop, the system evaluates $S(\mathcal{H}_t)$ against the three thresholds and applies the corresponding intervention from the pipeline described below.

\subsubsection{Three-Tier Context Capacity Envelope}

The context management pipeline is evaluated \emph{sequentially} at the beginning of each agent iteration, following a carefully designed ordering that ensures compression precedes restriction, and restriction precedes notification:

\paragraph{Tier 1: Force-Limit Compression ($S > C_{\text{force}}$).}
When the context payload exceeds $C_{\text{force}}$, the system initiates \emph{forced LLM-based semantic compression}. This is the most aggressive intervention: an auxiliary LLM pass summarizes the middle segment of the conversation history (see~\S\ref{sec:llm_compression}), replacing verbose tool I/O traces with a structured, information-preserving summary. If LLM compression continually fails (e.g., due to API errors), the system falls back to \emph{tail-drop compaction} (\S\ref{sec:tail_drop}), which retains the most recent messages within a target budget.

\paragraph{Tier 2: Hard-Limit Tool Restriction ($S > C_{\text{hard}}$).}
If the context exceeds $C_{\text{hard}}$, the system restricts the agent's action space to a minimal set of \emph{exit tools}: \texttt{phase\_done\_and\_summarize} (proactive phase transition) and \texttt{task\_done} (task completion). All other tools, including file operations, search, and code editing, are \emph{dynamically masked} from the LLM's tool schema for that iteration. A structured guard prompt is injected to inform the agent of the restriction and instruct it to either summarize progress or submit a final answer. This prevents the agent from generating tool calls that would further inflate the context while providing a clear escape path.

\paragraph{Tier 3: Soft-Limit Advisory Warning ($S > C_{\text{soft}}$).}
When the context exceeds $C_{\text{soft}}$ but remains below $C_{\text{hard}}$, a non-restrictive advisory warning is injected into the conversation. The agent retains full tool access but is prompted to proactively call \texttt{phase\_done\_and\_summarize} to preserve progress before harder limits are reached. This early-warning mechanism leverages the LLM's instruction-following capabilities to encourage self-regulated context management, reducing the frequency of forced compressions.

\paragraph{Ordering Rationale.}
The sequential evaluation order is critical: by applying compression before restriction, we maximize the probability that the agent can continue productive work after a force-limit trigger. Applying restriction after compression ensures that tool masking only activates when the context is genuinely intractable, avoiding premature capability reduction.

\subsubsection{Proactive Phase Transition: Agent-Initiated Context Reset}
\label{sec:proactive_phase}

The primary mechanism for \emph{proactive} context management is the \texttt{phase\_done\_and\_summarize} tool, which the agent can invoke at any point during execution. This tool implements a structured \emph{phase transition protocol} that performs the following operations atomically:

\begin{enumerate}[leftmargin=*]
    \item \textbf{Auto-Compression of Current Phase.} Before discarding the conversation history, the system extracts all messages from the first assistant response onward and passes them through the LLM-based compression pipeline (\S\ref{sec:llm_compression}) to generate a detailed, structured summary of the completed phase's work. This ensures that no critical information is lost during the transition.

    \item \textbf{Phase Resume Message Construction.} The system constructs a comprehensive \emph{phase resume} message that aggregates three information sources: (a)~the agent's self-authored \texttt{progress\_report} summarizing key findings and decisions; (b)~the agent's \texttt{remaining\_tasks} specifying concrete next steps; and (c)~the auto-compressed work log from the previous phase. Additionally, the system re-injects any preloaded skill protocols to maintain behavioral consistency across phases.

    \item \textbf{Context Reset and Reconstruction.} The conversation history is cleared entirely, and a fresh system prompt is rebuilt---including updated workspace state, followed by the phase resume message as the initial user input. The phase-local iteration counter is reset to zero while the \emph{global} iteration counter is preserved, preventing infinite phase cycling.

    \item \textbf{Trajectory Persistence.} A phase boundary event is emitted to the trajectory logging system, ensuring complete observability and enabling offline analysis of phase transition dynamics.
\end{enumerate}

This design follows the principle of \emph{structured forgetting}. Rather than retaining raw conversation traces indefinitely, the agent periodically distills its accumulated knowledge into a compact, high-fidelity summary and continues from a clean slate~\cite{ye2025agentfold}. The dual-counter mechanism (phase-local and global) provides both flexibility (resetting within-phase budgets) and safety (hard global budget), preventing degenerate behavior where the agent repeatedly transitions phases without making progress.

\subsubsection{LLM-Based Semantic Compression}
\label{sec:llm_compression}

The semantic compression module implements a three-stage pipeline that preserves the structural and informational integrity of the conversation while achieving significant size reduction:

\paragraph{Stage 1: Conversation Segmentation.}
The conversation history $\mathcal{H}$ is partitioned into three semantically meaningful segments:
\begin{itemize}[leftmargin=*]
    \item \textbf{Preamble} $\mathcal{P}$: All messages preceding the first assistant response (typically the system prompt and initial user task). These are preserved verbatim as they define the task specification and agent identity.
    \item \textbf{Compressible Segment} $\mathcal{C}$: Messages from the first assistant response up to (but not including) the last assistant response. This segment contains the bulk of tool interactions, intermediate reasoning, and accumulated observations that are candidates for compression.
    \item \textbf{Last Round} $\mathcal{L}$: The most recent assistant response and any subsequent tool/user messages. This segment is preserved verbatim to maintain the agent's immediate working context and avoid disrupting in-progress reasoning chains.
\end{itemize}
This segmentation ensures that compression targets the information-dense middle of the conversation while preserving both the foundational task context and the agent's active working state.

\paragraph{Stage 2: Structured Summarization.}
The compressible segment $\mathcal{C}$ is flattened into a linearized text representation, with tool result bodies truncated to a configurable maximum length to prevent individual oversized observations from dominating the compression input. This linearized representation is then passed to a dedicated LLM call guided by a structured compression prompt that mandates the following sections in the output summary:

\begin{itemize}[leftmargin=*]
    \item \emph{Task Understanding}: the agent's interpreted goal and any scope refinements;
    \item \emph{Completed Work}: chronological enumeration of actions with results, including exact file paths, function names, and data samples;
    \item \emph{Key Decisions \& Reasoning}: design choices, trade-offs, and hypothesis testing outcomes;
    \item \emph{Current Working State}: files modified, configurations set, and the agent's position in its workflow;
    \item \emph{Pending Work \& Plan}: remaining tasks with priority ordering;
    \item \emph{Errors, Blockers \& Workarounds}: failed approaches with root causes to prevent retry;
    \item \emph{Critical Data \& References}: exact identifiers, paths, and verbatim snippets needed for continuation.
\end{itemize}

This structured format is designed to maximize information density while providing the agent with clear organizational cues for retrieval during subsequent reasoning.

\paragraph{Stage 3: History Reconstruction.}
The compressed history is assembled as $\mathcal{H}' = \mathcal{P} \oplus [\text{compressed summary}] \oplus \mathcal{L}$, where the compressed summary is injected as a single user message annotated with a marker indicating the number of messages that were summarized. A subsequent system notification informs the agent that earlier raw tool traces have been absorbed into the summary, calibrating its expectations about available context.

\subsubsection{Tail-Drop Compaction}
\label{sec:tail_drop}

As a complementary fallback mechanism, tail-drop compaction provides a deterministic, non-LLM-dependent compression strategy. When invoked, it retains the system message (if present) and greedily accumulates messages from the \emph{most recent} backward, stopping when the cumulative size exceeds the target budget $B_{\text{target}} = 0.7 \times C_{\text{hard}}$. This ensures that the agent retains its most recent context---which is typically the most relevant for continued execution---while discarding older messages that are more likely to have been superseded by subsequent actions.

After compaction, a system notification is appended informing the agent that the context window was compacted and advising against re-reading large files unless necessary. This lightweight feedback loop helps the agent adapt its behavior to the reduced context availability.

\begin{table*}[t]
\centering
\caption{Context Management hyperparameters on LLM Backbones.}
\label{tab:LLM_backbone_context}
\small
\begin{tabular}{lccccc}
\toprule
\textbf{LLM Backbone} & \textbf{Context Limit} & \textbf{soft limit} $C_{\text{soft}}$ & \textbf{hard limit} $C_{\text{hard}}$ & \textbf{force limit} $C_{\text{force}}$ & \textbf{Tool Output Limit} \\
\midrule
GPT-5-Nano & 128K & 126K & 127K & 128K & 20K \\
Qwen3.6-35B-A3B & 262K & 260K & 261K & 262K & 20K \\
Gemini-3-Flash & 1M & 970K & 980K & 990K & 80K \\
\bottomrule
\end{tabular}
\end{table*}

\subsubsection{Auxiliary Context Guards}
\label{sec:aux_guards}

Beyond the three-tier capacity envelope, the framework incorporates several auxiliary guards that interact with the context management system:

\paragraph{No-Tool-Call Streak Detection.}
When the agent produces consecutive text-only responses without invoking any tool (indicating potential ``stuck'' behavior), an escalating sequence of nudge prompts is injected. After a configurable number of consecutive text-only turns $k_{\max}$, the system restricts the agent's tool set to \texttt{phase\_done\_and\_summarize} only, forcing a phase transition that resets the context and provides the agent with fresh instructions.

\paragraph{Phase-Local Iteration Budget.}
Near the end of each phase's iteration budget, the system proactively restricts the available tools to phase transition and task completion tools, preventing the agent from initiating new long-running operations that cannot be completed within the remaining budget.

\paragraph{Global Iteration Tail Guard.}
When the global iteration counter approaches the maximum, the system restricts tools to task completion only (\texttt{task\_done} or \texttt{finalize\_optimization}), ensuring that the agent submits a result rather than exhausting its entire compute budget without output.

\paragraph{Tool Output Truncation.}
Individual tool results exceeding a configurable character limit are truncated with tool-specific guidance messages (e.g., suggesting the use of line-range parameters for file reading or more selective search queries). This proactive truncation prevents single large tool observations from consuming a disproportionate share of the context budget and provides actionable feedback that helps the agent adopt more efficient information-gathering strategies in subsequent iterations.

\subsubsection{Implementation}

For the our experimental implementation, we keep fixed parameters for the context management mechanism across all four benchmarks as shown in~\Cref{tab:LLM_backbone_context}

\subsection{Discussion}
\label{sec:context_discussion}

Our context management framework differs from prior work in several key respects. First, the three-tier capacity envelope provides \emph{graduated} interventions rather than binary truncation, giving the agent maximum opportunity to self-manage its context before system-level overrides activate. Second, the proactive phase transition mechanism transforms context management from a purely system-side concern into a \emph{collaborative} protocol between the agent and the framework: the agent is encouraged (via soft-limit warnings and skill protocols) to strategically partition its workflow into phases, each with a clean context slate and a distilled summary of prior work. Third, the structured compression prompt ensures that the compression process is guided by task-relevant organizational principles, producing summaries that are directly actionable by the agent rather than generic text condensations.

Together, these mechanisms enable the agent to tackle complex, multi-step software engineering tasks that far exceed the native context window of the underlying LLM, while maintaining coherent state tracking and avoiding the information loss that characterizes simpler truncation-based approaches.

%% file: section/appendix/F_OptiMAS_design.tex
\section{OptiMAS Design}
\label{sec:appendix_optimas_design}

This section provides a comprehensive description of our proposed \textbf{OptiMAS} that drives the backward pass of our evolutionary training loop. The OptiMAS is itself an LLM-based agent that operates within the same ReAct framework as the evolutionary MAS it optimizes. It receives structured performance feedback from the forward pass (training batch diagnostics, validation signals, and full agent trajectories), analyzes failure patterns through a principled evidence-first methodology, and applies targeted interventions to the Multi-Agent System (MAS) configuration. A central design principle is \emph{domain agnosticism}: the OptiMAS's reasoning machinery, skill library, and intervention taxonomy are independent of the downstream task domain, enabling the same OptiMAS to drive evolution across code engineering, web research, tool orchestration, and other long-horizon agentic benchmarks.

\subsection{OptiMAS Architecture Overview}
\label{sec:opt_architecture}

The OptiMAS is essentially developed based on ReAct agent, inheriting all infrastructure capabilities, tool calling, context management, phase transitions, and skill loading, while introducing optimization-specific tools and a structured backward-pass workflow. At each training step, the OptiMAS receives:

\begin{itemize}[leftmargin=*]
    \item \textbf{Training batch diagnostics}: a structured report containing per-query outcomes (success/failure, scores), per-agent efficiency metrics (rounds, phases, tool call distributions, skill usage), and batch-level aggregates (accuracy, wall time).
    \item \textbf{Validation signal}: the current validation accuracy and its delta from the previous step, serving as an overfitting guard over the fixed validation set.
    \item \textbf{Previous optimization summary}: a textual record of the prior step's interventions, enabling continuity across optimization steps.
    \item \textbf{Full trajectory artifacts}: per-query directories containing structured audit logs (\texttt{tool\_calls.jsonl}), full ReAct conversation traces (per-agent, per-phase JSONL files), evaluation diagnostics, and workspace artifacts.
\end{itemize}

The OptiMAS processes this information with four typical workflows, \emph{Review \& Hypothesize}, \emph{Evidence Collection}, \emph{Intervention}, and \emph{Finalization}, guided by a hypothesis-driven long-term memory mechanism that facilitate cross-step experience condensation and ensures every modification is traceable and attributively grounded.

\subsection{Prompt Architecture}
\label{sec:opt_prompts}

The OptiMAS's prompt architecture comprises three interconnected components that collectively define its reasoning behavior.

\subsubsection{System Prompt}
\label{sec:opt_system_prompt}

The system prompt establishes the OptiMAS's identity, workflow structure, and decision-making principles. It begins by defining the OptiMAS's domain-agnostic role:

\begin{promptbox}
\small\texttt{You are \textbf{OptiMAS}, a general-purpose agent that improves Multi-Agent Systems (MAS). You are domain-agnostic: the MAS you optimize may target web research, code engineering, mathematical reasoning, GUI automation, multimodal analysis, or any other long-horizon task...
}
\end{promptbox}

The system prompt encodes the following critical components:

\paragraph{Hypothesis-Driven Workflow.}
The OptiMAS follows a structured six work modes: (1)~review optimization history and the persistent hypothesis ledger; (2)~gather evidence from trajectories using an audit-log-first discipline; (3)~formulate or update hypotheses with explicit observation, root cause, action, expected outcome structure; (4)~choose an intervention from the leveled action space; (5)~apply changes to the MAS configuration; (6)~validate and finalize.

\subsubsection{Backward Pass Task Prompt}
\label{sec:opt_backward_pass}

The backward pass template is instantiated at each optimization step with dynamically computed fields:

\begin{itemize}[leftmargin=*]
    \item \texttt{\{progress\}}: epoch and step counters indicating the current position in the training schedule.
    \item \texttt{\{validation\_metrics\}}: a formatted table presenting the current validation accuracy, accuracy delta.
    \item \texttt{\{prev\_summary\}}: the optimization summary from the previous step.
    \item \texttt{\{training\_batch\_diagnostics\}}: a comprehensive report including batch-level overview, per-query result tables (with scores, wall times, agent behavior summaries), and per-agent efficiency breakdowns (ReAct rounds, phases, tool call distributions, skill usage, delegation patterns).
    \item \texttt{\{artifact\_navigation\}}: workspace-relative paths to trajectory artifacts, evaluation diagnostics, and MAS configuration files.
\end{itemize}

\paragraph{Adaptive Sampling Awareness.}
The backward pass prompt explicitly informs the OptiMAS about the adaptive sampling strategy: some queries in each training batch are \emph{resampled} from the previous step's failures, enabling direct verification of whether the last optimization intervention was effective. The prompt instructs the OptiMAS to prioritize resampled failures as they represent ``strong direct evidence that your previous intervention was insufficient.''

\subsubsection{Initial Message Prompt}
\label{sec:opt_initial_message}

The initial message template provides the entry point for the OptiMAS's ReAct loop, incorporating the task description (backward pass prompt), preloaded skill protocols. This template uses the standard \texttt{\$\{task\}} placeholder mechanism shared with worker agents, ensuring consistency across the agent hierarchy.

\subsection{Skill Library}
\label{sec:opt_skills}

The OptiMAS employs a rich library of modular skills organized into three categories, \emph{system protocols}, \emph{preloaded protocols}, and \emph{optional skills}, following the same skill architecture used by worker agents but specialized for the optimization task.

\subsubsection{System Protocols (Always Active)}

These skills are inlined into the system prompt and define the OptiMAS's foundational operating contracts:

\paragraph{Handle-Workspace.}
Defines conventions for workspace navigation, file operations, and artifact management. Includes detailed guidance on using each available tool (file reading, writing, editing, searching, and document understanding) with OptiMAS-specific best practices.

\subsubsection{Preloaded Protocols (Persistent Across Phases)}

These skills are injected into the initial user message and re-injected after each phase transition, ensuring persistence across context resets:

\paragraph{Hypothesis Memory.}
The core knowledge management protocol. Defines the structure and lifecycle of optimization hypothesis long-term memory:

\begin{promptbox}
\#\#\# H<N> [status] (Step <created>)

- **Category**: positive / negative

- **Observation**: <specific behavior with evidence>

- **Root cause**: <why this happens>

- **Action**: <MAS change made or planned>

- **Expected outcome**: <measurable prediction>

- **Evidence**:

  - Step <X>: <observation>
\end{promptbox}

The memory enforces a rigorous status lifecycle, with explicit transition rules that require trajectory-level evidence for status changes. This prevents the OptiMAS from making unfounded generalizations from aggregate accuracy metrics alone. The hypothesis file persists on disk across all optimization steps, serving as the OptiMAS's long-term memory.

\paragraph{Phase Transition.}
Guides the OptiMAS's decisions about when to use \texttt{phase\_done\_and\_summarize} versus \texttt{task\_done}, with specific instructions for writing informative progress reports and persisting analysis findings to disk before phase boundaries.

\subsubsection{Optional Skills (Loaded On Demand)}

These skills are available for the OptiMAS to load when their specific expertise is needed:
\begin{itemize}
\item\textbf{Trajectory Analysis.} 
    The most comprehensive skill, providing a universal protocol for analyzing agent trajectories. 

\item\textbf{OptiMAS Skill Design.} 
    Provides the theory and practice of the three-tier skill system (system / preload / optional).

    \item\textbf{OptiMAS MAS Evolve.}
    Encodes the universal theory of MAS architecture evolution.

    \item\textbf{OptiMAS Create Sub-Agent.}
    A comprehensive recipe for adding sub-agents to the MAS.

    \item\textbf{OptiMAS Toolkits.} 
    Detailed guidance on the available tool suite~$\mathbb{T}$.

    \item\textbf{OptiMAS Investigator.} 
    Defines a read-only investigator sub-agent pattern for deep trajectory analysis on complex failing queries.
\end{itemize}

\subsection{Tool Suite}
\label{sec:opt_tools}

The OptiMAS has access to a curated set of tools that balance analytical capability with safety constraints.

\subsubsection{MCP Server Tools}

\begin{itemize}[leftmargin=*]
    \item \texttt{read\_file}: Read MAS configuration files, trajectory logs, and evaluation artifacts with optional line-range selection.
    \item \texttt{write\_file}: Create or modify MAS configuration files, prompts, skills, and working notes.
    \item \texttt{file\_edit}: Targeted line-level editing for surgical modifications to existing files.
    \item \texttt{grep\_file\_content}: Regular expression search across the workspace for pattern detection in trajectories and configurations.
    \item \texttt{list\_file\_tree}: Navigate the MAS configuration structure and trajectory artifact directories.
\end{itemize}

\subsubsection{Internal AI Tools}

\begin{itemize}[leftmargin=*]
    \item \texttt{text\_qa}: LLM-powered document understanding with sliding-window processing for large trajectory files. Used for grounded analysis of specific trajectory segments after the audit log has been reviewed directly.
    \item \texttt{image\_qa}: Vision-language model integration for analyzing visual artifacts (e.g., MAS crawled artifacts during rollout).
\end{itemize}

\subsubsection{Workflow Tools}

\begin{itemize}[leftmargin=*]
    \item \texttt{load\_skill}: Dynamically load optional skills on demand, enabling the OptiMAS to access specialized knowledge only when needed.
    \item \texttt{phase\_done\_and\_summarize}: Trigger a phase transition with auto-compression, enabling the OptiMAS to work on complex optimization tasks that exceed a single context window.
    \item \texttt{task\_done}: Mark the optimization step as complete.
\end{itemize}

\subsubsection{Optimization-Specific Tool: \texttt{finalize\_optimization}}

The \texttt{finalize\_optimization} tool is the OptiMAS's delivery mechanism. Upon invocation, it executes a multi-stage validation pipeline before accepting the optimization step. The programmatic completeness check of Plan oriented short-memory is implemented in this stage. If any stage fails, detailed error messages are returned to the OptiMAS, which can fix the issues and re-invoke \texttt{finalize\_optimization}. This iterative validation loop ensures that every delivered optimization step produces a structurally valid, deployable MAS configuration and the OptiMAS working consistently without forgetting and hallucination as well.

\subsection{Hypothesis-Driven Long-Term Memory}
\label{sec:opt_hypothesis_memory}

A distinguishing feature of our OptiMAS is its \emph{hypothesis-driven long-term memory} system, implemented through two persistent artifacts:

\paragraph{Hypothesis.}
The data-based structural hypothesis persists across all optimization steps and epochs, serving as the OptiMAS's cumulative knowledge base. Each hypothesis records an observation, root cause analysis, chosen action, concrete intervention, expected outcome, and evolving evidence trail. Hypotheses cannot be confirmed without trajectory-level evidence, and cannot be refuted based on aggregate accuracy metrics alone.

\subsection{Domain Adaptability}
\label{sec:opt_domain}

The OptiMAS's domain-agnostic design is achieved through a strict separation of concerns:

\begin{itemize}[leftmargin=*]
    \item The \emph{universal optimization machinery}: Hypothesis management, trajectory analysis protocols, failure taxonomy, action space, skill design theory, MAS architecture evolution, are all encoded in domain-agnostic skills and prompts.
    \item The \emph{backward pass template} dynamically injects evaluation guidance based on the benchmark type detected from the batch report, without modifying the universal template structure.
\end{itemize}

This architecture enables the same OptiMAS configuration to drive evolution across heterogeneous benchmarks, from code engineering to web research to long reasoning too orchestration, with the core optimization methodology remains invariant.

%% file: section/appendix/G_case_study.tex
\section{Case Study}

\input{section/figure/swebench_case}

\noindent\textbf{Evolutionary MAS on SWE-Bench-Verified.} 
To investigate the macro-level behavioral optimization of the evolving MAS, we trace the structural mutations captured in Figure~\ref{fig:swebench_mas_evolution} at the final evolution ($\mathcal{E}^*$). The results demonstrates that the OptiMAS transform single agent in favor of granular role specialization. Specifically, the OptiMAS instantiates an architectural core on the \emph{Worker} agent, expanding its repository to eight highly specialized meta-skills, including \emph{logic preservation audit} and \emph{forced delegation gates}. Concurrently, resource-intensive sandbox tracking and post-patch validation are decoupled into localized execution loops managed by the \emph{Refiner}, \emph{Reproducer} and \emph{Verifier} sub-agents, respectively. Rather than relying on fragile localized prompt engineering, these emergent architectural workflows, distinct sub-agent delegation cascades, and targeted skill discovery runs validate that our evolutionary framework systematically acquires the structural rigor required to navigate complex code repositories.

%% file: section/figure/swebench_case.tex
\begin{figure*}
    \centering
    \centerline{\includegraphics[width=0.85\textwidth]{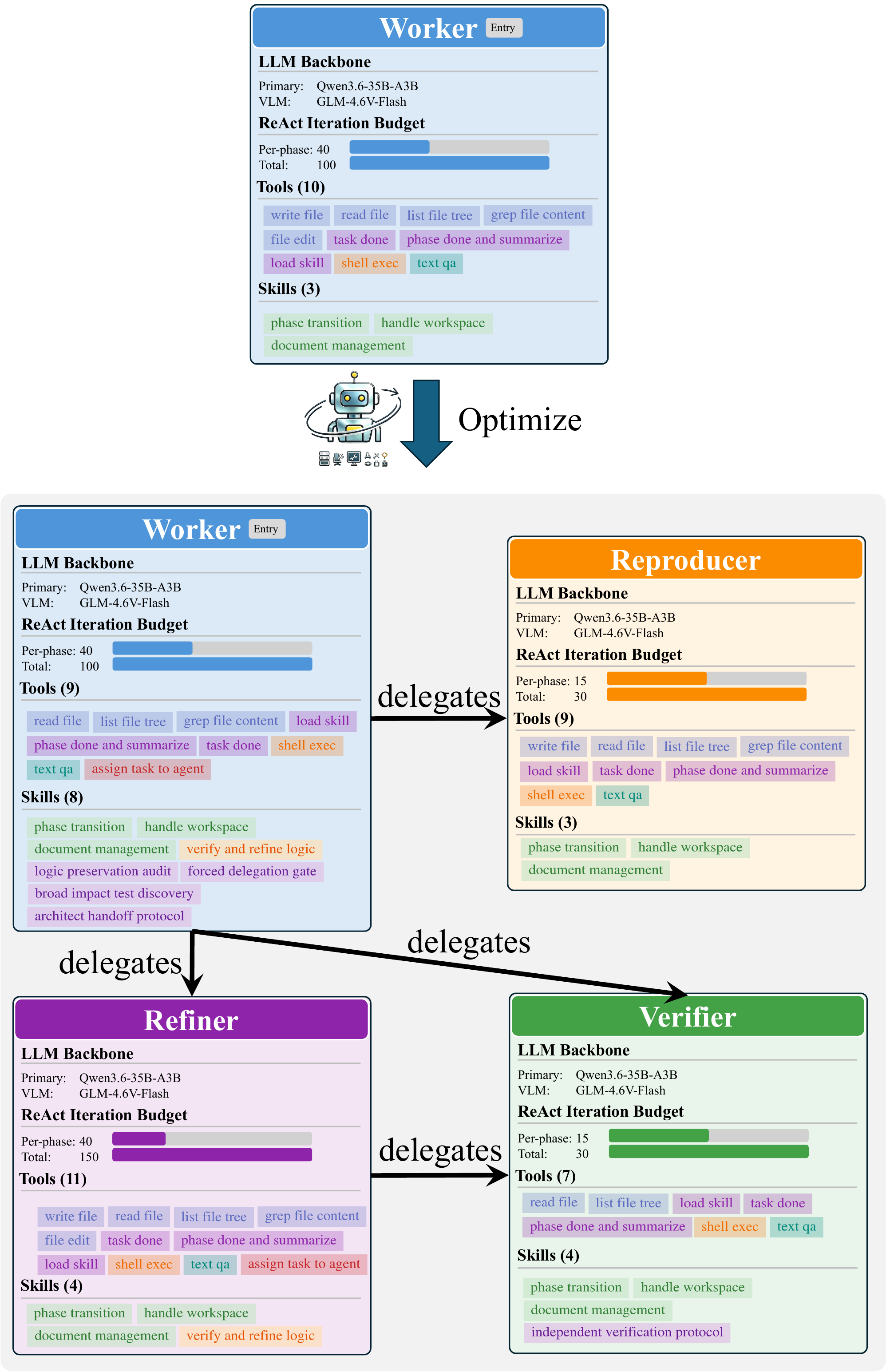}}
    \caption{\textbf{Architectural Topography of the Evolved Multi-Agent System (MAS) on SWE-Bench-Verified.} This schematic contrasts the initial configuration $\mathcal{E_0}$ against the evolved architecture ($\mathcal{E^*}$). Through our OptiMAS optimization, the system undergoes severe structural transformation: transitioning from a standard flat-worker scheme into a modular, hierarchical coalition composed of specialized agents (\emph{Worker}, \emph{Reproducer}, \emph{Refiner}, and \emph{Verifier}). Each sub-agent customized ReAct iteration budgets, and differentiated, priority-driven tool and skill allocations.}
    \label{fig:swebench_mas_evolution}
\end{figure*}